\documentclass[aps,english,preprint,nofootinbib]{revtex4}

\usepackage{graphicx}
\usepackage{hyperref}
\usepackage{amsmath, amssymb}
\usepackage{babel}
\usepackage{color}
\usepackage{mathrsfs}
\usepackage{slashed}

\begin{document}

\title{Improved $K_{e3}$ radiative corrections sharpen the $K_{\mu 2}$--$K_{l3}$ discrepancy}

\author{Chien-Yeah Seng$^{1}$}
\author{Daniel Galviz$^{1}$}
\author{Mikhail Gorchtein$^{2,3,4}$}
\author{Ulf-G. Mei{\ss}ner$^{1,5,6}$}

\affiliation{$^{1}$Helmholtz-Institut f\"{u}r Strahlen- und Kernphysik and Bethe Center for
  Theoretical Physics,\\ Universit\"{a}t Bonn, 53115 Bonn, Germany}
\affiliation{$^{2}$Helmholtz Institute Mainz, D-55099 Mainz, Germany}
\affiliation{$^{3}$GSI Helmholtzzentrum f\"ur Schwerionenforschung, 64291 Darmstadt, Germany}
\affiliation{$^{4}$Johannes Gutenberg University, D-55099 Mainz, Germany}
\affiliation{$^{5}$Institute for Advanced Simulation, Institut f\"ur Kernphysik and J\"ulich Center
  for Hadron Physics, Forschungszentrum J\"ulich, 52425 J\"ulich, Germany}
\affiliation{$^{6}$Tbilisi State  University,  0186 Tbilisi, Georgia}

\date{\today}

\begin{abstract}
The measurements of $V_{us}$ in leptonic $(K_{\mu 2})$ and semileptonic $(K_{l3})$ kaon decays exhibit
a $3\sigma$ disagreement, which could originate either from physics beyond the Standard Model or
some large unidentified Standard Model systematic effects. Clarifying this issue requires a careful
examination of all existing Standard Model inputs. Making use of a newly-proposed computational
framework and the most recent lattice QCD results, we perform a comprehensive re-analysis of the
electroweak radiative corrections to the $K_{e3}$ decay rates that achieves an unprecedented level of
precision of $10^{-4}$, which improves the current best results by almost an order of magnitude. No
large systematic effects are found, which suggests that the electroweak radiative corrections
should be removed from the ``list of culprits'' responsible for the $K_{\mu 2}$--$K_{l3}$ discrepancy. 

\end{abstract}

\maketitle


\section{Introduction}

Despite the discovery of the Higgs boson in the year 2012~\cite{Aad:2012tfa,Chatrchyan:2012ufa} that
completed the particle spectrum in the Standard Model (SM), there exist numerous observed phenomena
in astrophysics,  e.g. dark matter, dark energy and the matter-antimatter asymmetry,
that do not find their explanations within this theory framework and thus call for physics beyond
the Standard Model (BSM). Unfortunately, all direct searches in high-energy colliders have so far
returned null results. On the other hand, precision experiments have observed several interesting
anomalies in flavor physics that point towards the possible existence of BSM physics. This
research concerns one of these observed anomalies, namely the irregularities in the top-row
Cabibbo-Kobayashi-Maskawa (CKM) matrix elements.

The unitarity of the CKM matrix is a rigorous SM prediction~\cite{Cabibbo:1963yz,Kobayashi:1973fv}.
In particular, the top-row CKM unitarity (which is also known as the Cabibbo unitarity) that involves the matrix elements $V_{ud}$ and $V_{us}$
($V_{ub}$ is negligible) has received the most attention because they can be measured to high
precision in hadron and nuclear beta decays. Recently, a series of improvements in the
theory~\cite{Seng:2018yzq,Seng:2018qru,Gorchtein:2018fxl,Czarnecki:2019mwq} of the electroweak
radiative corrections (RC) in the extraction of $V_{ud}$ led to an apparent deviation of the
Cabibbo unitarity at a level of $3\sigma$~\cite{Zyla:2020zbs}. However, in this work we will
not focus on $V_{ud}$, but rather on $V_{us}$ which possesses yet another interesting anomaly by itself.

Let us focus on the two best determinations of the matrix element $V_{us}$, which come from leptonic ($K_{l2}$) and semileptonic ($K_{l3}$) kaon
decays respectively. From the leptonic kaon {\color{black} and pion} decay, the following ratio is obtained:
\begin{equation}
\frac{|V_{us}|f_{K^+}}{|V_{ud}|f_{\pi^+}}=0.23871(20)\left[\frac{\Gamma_{K\rightarrow \mu\nu(\gamma)}}
{\Gamma_{\pi\rightarrow\mu\nu(\gamma)}}\right]^{\frac{1}{2}}~,\label{eq:Kl2ratio}
\end{equation} 
where $f_{K^+}$ and $f_{\pi^+}$ are the $K^+$ and the $\pi^+$ decay constant, respectively, which
require lattice QCD inputs. The theory uncertainty on the right-hand side is less than $10^{-3}$,
thanks to the cancellation of the common electroweak RC to the leptonic kaon and pion decay
rate~\cite{Marciano:2004uf,Cirigliano:2011tm}. {\color{black} Combining this expression} with the $N_f=2+1+1$ FLAG average of
$f_{K^+}/f_{\pi^+}$~\cite{FlavourLatticeAveragingGroup:2019iem} and the recent value of $V_{ud}$ obtained from superallowed
beta decays~\cite{Seng:2018yzq}, the following result is quoted in PDG~2020~\cite{Zyla:2020zbs}:
\begin{equation}
|V_{us}|=0.2252(5)\:\:({\color{black}K_{\mu 2}/\pi_{\mu 2} + \mathrm{superallowed}})\label{eq:VusKl2}
\end{equation}
Meanwhile, in the semileptonic kaon decay process $K\to\pi l^+\nu(\gamma)$ one does not measure a
ratio, but obtains $V_{us}$ directly from the decay rate, where the SM inputs include the electroweak
RC, the $K\pi$ form factors and the SU(2) isospin-breaking effects (we postpone the detailed
discussions to the main text). With the most recent theory inputs of these quantities, PDG~2020
quotes the following result:
\begin{equation}
|V_{us}|=0.2231(4)_\mathrm{exp+RC}(6)_\mathrm{lat}\:\:(K_{l3})\label{eq:VusKl3}
\end{equation}
We observe a $\sim 3\sigma$ disagreement between the numbers in Eq.\eqref{eq:VusKl2} and
\eqref{eq:VusKl3}, with a $\sim 1\%$ difference between the two central values. This provides
another interesting hint to the existence of BSM physics~\cite{Belfatto:2019swo,Tan:2019yqp,Grossman:2019bzp,Coutinho:2019aiy,Cheung:2020vqm,Crivellin:2020lzu,Endo:2020tkb,Capdevila:2020rrl,Kirk:2020wdk,Crivellin:2020oup} which,
to some extent, is even more promising than the top-row CKM unitarity deficit. In fact, the extraction
of $V_{us}$ is free from complicated nuclear-structure uncertainties (except those that enter $V_{ud}$
in Eq.~\eqref{eq:Kl2ratio}, whose effect on $V_{us}$ is subdominant to the existing uncertainties).
For instance, if the total uncertainty in Eqs.\eqref{eq:VusKl2} and \eqref{eq:VusKl3} is reduced
to $4\times 10^{-4}$ or below, with the central values unchanged, the discrepancy will reach
$5\sigma$ which is sufficient to claim an observation of a BSM signal.
Achieving this final goal requires a careful re-analysis of all the SM inputs, not just to reduce
their uncertainties but also to make sure that no large unidentified SM corrections were missed in existing analyses. 

In this work, we study a particularly important SM correction to the kaon semileptonic decay,
namely the electroweak RC. Earlier studies of this topic by Ginsberg~\cite{Ginsberg:1966zz,Ginsberg:1968pz,Ginsberg:1969jh,Ginsberg:1970vy}, Becherrawy~\cite{Becherrawy:1970ah} and later by Bytev \textit{et al.}~\cite{Bytev:2002nx} and Andre~\cite{Andre:2004tk} assumed specific models for the strong and electroweak interactions which made a rigorous analysis of the theory uncertainties rather challenging. Another class of works,
e.g. by Garc\'{\i}a and Maya~\cite{Garcia:1981it} and by Ju\'{a}rez-Le\'{o}n \textit{et al.}~\cite{JuarezLeon:2010tj,Torres:2012ge,Neri:2015eba} put more emphasis on the so-called ``model-independent'' piece in the
long-distance electromagnetic corrections (i.e. the convection term contribution, which we will explain
in the main text) but were unable to place any constrain on the ``model-dependent'' piece originating
from non-perturbative Quantum Chromodynamics (QCD) at the hadronic scale. So far, the only approach
that allows a systematic error analysis in every part of the electroweak RC has been the chiral
perturbation theory (ChPT) calculation by Cirigliano \textit{et al.}~\cite{Cirigliano:2001mk,Cirigliano:2004pv,Cirigliano:2008wn}, where the most general electroweak interactions between hadrons and dynamical
photons~\cite{Urech:1994hd} and leptons~\cite{Knecht:1999ag} are arranged according to increasing
powers of $p/\Lambda_\chi$, where $p$ is a typical small momentum scale in such interactions and
$\Lambda_\chi \simeq 4\pi F_\pi$ is the chiral symmetry breaking scale, with $F_\pi=92.1$~MeV the
pion decay constant. Within this framework, the long-distance electromagnetic RC to $K_{l3}$ decay
is calculated to $\mathcal{O}(e^2p^2)$, and the theory uncertainty comes from two major sources: The
unknown low-energy constants (LECs) at $\mathcal{O}(e^2p^2)$, and the neglected contributions of the
order $\mathcal{O}(e^2p^4)$. Both uncertainties are estimated to be of the order $10^{-3}$. At this
point it seems formidable to make any further progress within the same theory framework, because (1)
the LECs are only calculable within phenomenological
models~\cite{Ananthanarayan:2004qk,DescotesGenon:2005pw} with outcomes that are highly uncertain,
and (2) to reduce the higher-order corrections one needs to perform a full two-loop ChPT calculation
which is not only technically challenging but more importantly, involves even more unknown LECs.  

A series of preparatory works were done since early 2020 in order to eventually overcome
the difficulties mentioned above. First, a new theory framework based on the hybridization of
the classical Sirlin's approach~\cite{Sirlin:1977sv,Seng:2021syx} and modern ChPT
was formulated~\cite{Seng:2019lxf} in order to resum the most important
$\mathcal{O}(e^2p^{2n})$ effects while retaining the full model-independent characteristics in the
traditional ChPT approach. Next, lattice QCD was introduced to study the part of the RC in semileptonic
decays that carries the largest hadronic uncertainties, namely the axial $\gamma W$-box diagram. The first
calculation was done on the pion~\cite{Feng:2020zdc}, which removed the dominant theory uncertainty in
the semileptonic pion decay and also confirmed the result of the previous dispersion-relation analysis
of the RC in free neutron~\cite{Seng:2020wjq}. Shortly after that, following the suggestion in
Ref.\cite{Seng:2020jtz} a new lattice calculation of the $K\pi$ axial $\gamma W$-box in the
flavor SU(3) limit was performed~\cite{Ma:2021azh}. Up to this point, we finally have all the necessary ingredients
and are in the position to present a fully-updated numerical analysis of the electroweak RC in kaon
semileptonic decays that eventually reduces the existing theory uncertainty by almost an order of
magnitude, i.e. to the level of $10^{-4}$. 

The main results in this study were presented in an earlier paper~\cite{Seng:2021boy}, and here we will show all the details. We concentrate on the $K_{e3}$ channel and not $K_{\mu 3}$ throughout this study for reasons that will
become clear in the main text. The contents of this work are arranged as follows. In
Section~\ref{sec:basic} we introduce the basic notation and set up our theory framework.
In Sections~\ref{sec:ana}--\ref{sec:axialbox} we present our update of the contributions from
the ``virtual'' electroweak RC; in particular, we demonstrate in Section~\ref{sec:axialbox} how
the most recent lattice QCD results are used to constrain the hadronic uncertainties in the
physical $K\pi$ axial $\gamma W$-box diagram. The contribution from the real-photon emission
process is calculated in Section~\ref{sec:brem}. In Section~\ref{sec:comparison} we discuss how
our new results should be interpreted in the ChPT language, and show the numerical improvement
against the existing calculations. Final discussions and conclusions are provided in
Section~\ref{sec:final}.

\section{\label{sec:basic}Notation and setup}

One of the most important avenues to extract $V_{us}$ is the inclusive kaon semileptonic decay $K_{l3}$,
i.e. the process $K(p)\to \pi(p')+ l^+(p_l)+\nu_e(p_\nu)+n\gamma$, where $l=e,\mu$, and $n\geq 0$ is
the number of photons in the final state. It will be evident later that the case $l=e$ allows for a
much better control of the theory uncertainties, so throughout this paper, we will concentrate on
this particular case. If all massless final-state particles are left unobserved, the
differential decay rate of the process is fully described by three independent, dimensionless
Lorentz-invariant variables\footnote{In the existing literature $x$ is more often defined as $P^2$,
which carries a dimension.}:
\begin{equation}
x\equiv\frac{P^2}{M_K^2},\:\:y\equiv\frac{2p\cdot p_e}{M_K^2},\:\:z\equiv\frac{2p\cdot p'}{M_K^2}~,
\label{eq:xyz}
\end{equation}
where $P\equiv p-p'-p_e$. Notice that $x$ is strictly zero (neglecting neutrino mass)
for $n=0$, but may take a non-zero value when $n\geq 1$. We may have as well introduced the usual Mandelstam
variables $s\equiv(p'+p_e)^2$, $t\equiv(p-p')^2$ and $u\equiv(p-p_e)^2$, but none of them is
independent of $\{x,y,z\}$.

At $\mathcal{O}(G_F^2)$ (where $G_F=1.1663787(6)\times 10^{-5}$~GeV$^{-2}$ is the Fermi constant
extracted from muon decay~\cite{MuLan:2012sih}), only the $n=0$ process contributes to
the $K_{e3}$ decay rate. Its corresponding tree-level amplitude is given by:
\begin{equation}
M_0=-\frac{G_F}{\sqrt{2}}\bar{u}_\nu\gamma^\mu(1-\gamma_5)v_e F^{K\pi}_\mu(p',p)~,
\end{equation} 
where the effects of the strong interaction are fully contained in the following hadronic
matrix element of the charged weak current:
\begin{equation}
F_\mu^{K\pi}(p',p)\equiv\left\langle \pi(p')\right|(J^W_\mu)^\dagger\left|K(p)\right\rangle =
V_{us}^*\left[f_+^{K\pi}(t)(p+p')_\mu+f_-^{K\pi}(t)(p-p')_\mu\right]~.
\end{equation}
The equation above defines the charged weak form factors $f_\pm^{K\pi}(t)$\footnote{We wish to remind
the readers that our sign convention for the form factors is $f_+^{K\pi}(0)<0$, which is also adopted
in our previous works, e.g.\cite{Seng:2019lxf,Seng:2020jtz}, but may be opposite to other existing
literature. This serves to be consistent with the sign convention of the charged weak current
$(J_W^\mu)^\dagger$ derived from ChPT.}. It is also customary to define a third form factor:
\begin{equation}
f_0^{K\pi}(t)\equiv f_+^{K\pi}(t)+\frac{t}{M_K^2-M_\pi^2}f_-^{K\pi}(t)~,
\end{equation}
and call $f_+^{K\pi}(t)$ and $f_0^{K\pi}(t)$ the ``vector'' and ``scalar'' form factor, respectively.
From the definition above, it is obvious that $f_0^{K\pi}(0)=f_+^{K\pi}(0)$, so another common
step is to factor out their $t=0$ value:
\begin{equation}
\bar{f}_{+,0}(t)\equiv\frac{f_{+,0}^{K\pi}(t)}{f_+^{K\pi}(0)}~.
\end{equation} 
There are several different ways to parameterize $\bar{f}_{+,0}(t)$, e.g. Taylor expansion,
monopole parameterization and dispersive parameterization. The interested reader may consult
Ref.~\cite{Lazzeroni:2018glh} and references therein for the details, and we will also
come back to this point in Section~\ref{sec:Born}.

It is instructive to display explicitly the absolute square of the tree-level amplitude above
(upon summing over the lepton spin, as we will always do throughout this work):
\begin{equation}
|M_0|^2(x,y,z)=G_F^2F^{K\pi}_\mu(p',p)(F_\nu^{K\pi}(p',p))^*\mathrm{Tr}
\left[\slashed{P}\gamma^\mu(\slashed{p}_e-m_e)\gamma^\nu(1-\gamma_5)\right]~.\label{eq:M20}
\end{equation}
Here we purposely retain the $x$-dependence in the formula above despite the fact that $x=0$ when
$n=0$. The $x$-dependence becomes important later when we discuss the squared amplitude of the
bremsstrahlung process. The impact of the form factors $f_\pm^{K\pi}$ on the tree-level decay rate
relies heavily on the leptonic trace in Eq.\eqref{eq:M20}. Suppose we define:
\begin{equation} 
H(a,b)\equiv\mathrm{Tr}\left[\slashed{P}(\slashed{p}+a\slashed{p}')(\slashed{p}_e-m_e)
(\slashed{p}+b\slashed{p}')(1-\gamma_5)\right]_{x=0}~,
\end{equation}
then a straightforward calculation shows:
\begin{eqnarray}
H(+1,+1)&=&-2M_K^4\left[4(y-1)(y+z-1)+4r_\pi-r_e(r_\pi+4y+3z-3)+r_e^2\right]\nonumber\\
H(+1,-1)&=&H(-1,+1)\nonumber\\
&=&-2M_K^4r_e\left[-r_e+r_\pi+2y+z-3\right]\nonumber\\
H(-1,-1)&=&-2M_K^4r_e\left[r_e-r_\pi+z-1\right]~,
\end{eqnarray}
where $r_\pi\equiv M_\pi^2/M_K^2$ and $r_e\equiv m_e^2/M_K^2$. We observe that only $H(+1,+1)$ is
not explicitly suppressed by the factor $r_e\approx 10^{-6}$. Following the notations in
Appendix~\ref{sec:PS}, the decay rate at $\mathcal{O}(G_F^2)$ in given by:
\begin{equation}
\left(\Gamma_{K_{e3}}\right)_\mathrm{tree}=\frac{M_K}{256\pi^3}\int_{\mathcal{D}_3}dydz|M_0|^2(0,y,z)~.
\end{equation}
From the argument above, it is apparent that only $f_+^{K\pi}(t)$, and not $f_-^{K\pi}(t)$, is
relevant in $\left(\Gamma_{K_{e3}}\right)_\mathrm{tree}$. Of course the actual value of
$\left(\Gamma_{K_{e3}}\right)_\mathrm{tree}$ depends on the specific parameterization of $\bar{f}_+(t)$
and the parameters therein, but the impact of the different choices is generically of the
order $0.1\%$. Since in this paper $\left(\Gamma_{K_{e3}}\right)_\mathrm{tree}$ serves only as
a normalization factor to the already-small RC, such a difference is completely negligible.

The electroweak RC induces a shift of the tree-level decay rate:
$\left(\Gamma_{K_{e3}}\right)_\mathrm{tree}\rightarrow\left(\Gamma_{K_{e3}}\right)_\mathrm{tree}
+\delta \Gamma_{K_{e3}}$. We define the quantity:
\begin{equation}
\delta_{K_{e3}}\equiv\frac{\delta\Gamma_{K_{e3}}}{\left(\Gamma_{K_{e3}}\right)_\mathrm{tree}}
\end{equation}
that represents the fractional correction to the decay rate, and we will discuss its relation to
the different quantities within the ChPT framework in Section~\ref{sec:comparison}. To match the
precision level of current and near-future experiments, we need a theoretical prediction of
$\delta_{K_{e3}}$ up to $\mathcal{O}(\alpha)$. At this level, the only two contributors are (1)
the $\mathcal{O}(G_F\alpha)$ electroweak RC to the $n=0$ decay amplitude, and (2) the tree-level
contribution from the $n=1$ process. We will spend the next few sections discussing these two contributions.

\section{\label{sec:ana}Virtual correction: analytic pieces}

We start by discussing the virtual corrections, i.e. the $\mathcal{O}(G_F\alpha)$ electroweak RC
to the $n=0$ decay amplitude. It is possible to express such corrections entirely in terms of
perturbations to the charged weak form factors, i.e\footnote{Using the on-shell condition, one can
show that other leptonic bilinear structures, such as $i\epsilon^{\lambda\mu\nu\alpha}p_\mu p'_\nu
p_{e\alpha}\bar{u}_\nu\gamma_\lambda(1-\gamma_5)v_e$, are linear combinations of $\bar{u}_\nu(\slashed{p}
\pm\slashed{p}')(1-\gamma_5)v_e$.  },
\begin{equation}
\delta M_\mathrm{vir}=-\frac{G_F}{\sqrt{2}}V_{us}^*\bar{u}_\nu\gamma^\mu(1-\gamma_5)v_e\left[(p+p')_\mu \delta f_+^{K\pi}+(p-p')_\mu\delta f_-^{K\pi}\right]~.
\end{equation}
The only complication is that $\delta f_\pm^{K\pi}$ are complex functions of two variables, e.g. $\{y,z\}$, 
rather than real functions of a single variable $t$.

The virtual contribution to $\delta_{K_{e3}}$ at $\mathcal{O}(\alpha)$ arises from the interference
between $M_0$ and $\delta M_\mathrm{vir}$, i.e. $|M_0|^2\to |M_0|^2+\delta|M|^2_\mathrm{vir}$, with
$\delta |M|^2_\mathrm{vir}\equiv2\mathfrak{Re}\left\{M_0^*\delta M_\mathrm{vir}\right\}$. Again, by
restricting ourselves to $K_{e3}$, we only need to know $\delta f_+^{K\pi}$ in order to determine
the perturbation to the $n=0$ squared amplitude:
\begin{equation}
\delta |M|^2_\mathrm{vir}(y,z)=2|M_0|^2(0,y,z)\frac{\mathfrak{Re}\left\{\delta f_+^{K\pi}\right\}}{f_+^{K\pi}(t)}+\mathcal{O}(r_e)~.
\end{equation} 
Based on the theory framework outlined in Refs.\cite{Seng:2019lxf,Seng:2020jtz}, the
$\mathcal{O}(G_F\alpha)$ virtual corrections to the $n=0$ decay amplitude can be summarized by
the following equation:
\begin{eqnarray}
\delta M_\mathrm{vir}&=&\left[-\frac{\alpha}{2\pi}\left(\ln\frac{M_W^2}{M_Z^2}
+\frac{1}{4}\ln\frac{M_W^2}{m_e^2}
-\frac{1}{2}\ln\frac{m_e^2}{M_\gamma^2}+\frac{9}{8}+\frac{3}{4}a_\mathrm{pQCD}\right)+\frac{1}{2}
\delta_{\mathrm{HO}}^\mathrm{QED}\right]M_0\nonumber\\
&&+\delta M_2+\delta M_3+\delta M_{\gamma W}~.\label{eq:deltaMEW}
\end{eqnarray}
Let us briefly explain the notation above,  all the details are given in Ref.\cite{Seng:2019lxf}.
First, the terms in the square bracket come from the ``weak'' RC including its $\mathcal{O}(\alpha_s)$
perturbative QCD (pQCD) corrections $a_\mathrm{pQCD}\approx 0.068$, the electron wavefunction
renormalization, and the resummation of the large QED logs represented by
$\mathrm{\delta}_\mathrm{HO}^\mathrm{QED}=0.0010(3)$~\cite{Erler:2002mv}. An infinitesimal photon
mass $M_\gamma$ is introduced to regularize the infrared (IR) divergence in the electron
wavefunction renormalization. Next, the quantities $\delta M_{2,3}$
represent the contributions from two separate pieces of the electromagnetic RC to the charged weak
form factors, known as the ``two-point function'' and ``three-point function'', respectively.
Finally, $\delta M_{\gamma W}$ represents the contribution from the $\gamma W$-box diagram:
\begin{equation}
\delta M_{\gamma W}=-\frac{G_Fe^2}{\sqrt{2}}\int\frac{d^4q'}{(2\pi)^4}\frac{M_W^2}{M_W^2-q^{\prime 2}}
\frac{\bar{u}_\nu\gamma^\nu(\slashed{q}'\gamma^\mu-2p_e^\mu)(1-\gamma_5)v_e}{\left [(p_e-q')^2-m_e^2\right]
\left[q^{\prime 2}-M_\gamma^2\right]}T_{\mu\nu}^{K\pi}(q';p',p)~,
\end{equation}
where we have introduced the so-called ``generalized Compton tensor'' $T_{\mu\nu}^{K\pi}$ which plays
a central role in the upcoming analysis:
\begin{equation}
T^{\mu\nu}_{K\pi}(q';p',p)\equiv\int d^4x \, e^{iq'\cdot x}\left\langle \pi(p')\right|T\{J_\mathrm{em}^\mu(x)
J_W^{\nu\dagger}(0)\}
\left|K(p)\right\rangle~,
\end{equation}
where $T\{\ldots\}$ denotes the conventional time-ordering.
It satisfies the following Ward identities:
\begin{eqnarray}
q'_\mu T_{K\pi}^{\mu\nu}(q';p',p)&=&-iF_{K\pi}^\nu(p',p)\nonumber\\
q_\nu T_{K\pi}^{\mu\nu}(q';p',p)&=&-iF_{K\pi}^\mu(p',p)-i\Gamma_{K\pi}^\mu(q';p',p)~,\label{eq:Ward}
\end{eqnarray}
with $q\equiv p'+q'-p$, and 
\begin{equation}
\Gamma^{\mu}_{K\pi}(q';p',p)\equiv\int d^4x \, e^{iq'\cdot x}\left\langle \pi(p')\right|
T\{J_\mathrm{em}^\mu(x)\partial\cdot
J_W^{\dagger}(0)\}\left|K(p)\right\rangle~.\label{eq:Gammamu}
\end{equation}
The first line in Eq.\eqref{eq:Ward} is a consequence of the exact conservation of the
electromagnetic current, while the second line entails the partial conservation of the
charged weak current. Expressing hadronic matrix elements in terms of integrals with respect to $T_{\mu\nu}$ is a classical technique in hadron physics that appears also in, e.g., the Cottingham's approach to the hadronic mass splittings~\cite{Cottingham:1963zz,Gasser:1974wd,Bardeen:1988zw,Walker-Loud:2012ift,Gasser:2015dwa,Gasser:2020mzy}.

Using now the following Dirac matrix identity:
\begin{equation}
\gamma^\mu\gamma^\nu\gamma^\alpha=g^{\mu\nu}\gamma^\alpha-g^{\mu\alpha}\gamma^\nu+g^{\nu\alpha}\gamma^\mu
-i\epsilon^{\mu\nu\alpha\beta}\gamma_\beta\gamma_5~,
\end{equation}
(with $\epsilon^{0123}=-1$) one splits the $\gamma W$-box diagram into two pieces: $\delta M_{\gamma W} =
\delta M_{\gamma W}^a+\delta M_{\gamma W}^b$, where the antisymmetric tensor is contained in the
second term\footnote{We used to label them as $\delta M_{\gamma W}^V$ and $\delta M_{\gamma W}^A$ in
Ref.\cite{Seng:2020jtz}, but this may cause confusions with notations of box diagrams in some literature
when we further divide the contributions from the vector and axial charged weak current in
$T_{\mu\nu}^{K\pi}$, so here we adopt an alternative labeling.}. A great simplification is observed
upon combining $\delta M_2$ with $\delta M_{\gamma W}^a$~\cite{Seng:2020jtz}:
\begin{eqnarray}
\delta M_2+\delta M_{\gamma W}^a&=&\frac{\alpha}{2\pi}\left[\ln\frac{M_W^2}{m_e^2}+\frac{3}{4}+\frac{1}{2}
\tilde{a}_g^\mathrm{res}\right]M_0+\frac{G_Fe^2}{\sqrt{2}}\bar{u}_\nu\gamma_\lambda(1-\gamma_5)v_e
\int\frac{d^4q'}{(2\pi)^4}\frac{M_W^2}{M_W^2-q^{\prime 2}}\nonumber\\
&&\times\frac{1}{(p_e-q')^2-m_e^2}\left\{\frac{2p_e\cdot q'q^{\prime\lambda}}{(q^{\prime 2}-M_\gamma^2)^2}
T^\mu_{K\pi\mu}(q';p',p)+\frac{2p_{e\mu}}{q^{\prime 2}-M_\gamma^2}T^{\mu\lambda}_{K\pi}(q';p',p)\right.\nonumber\\
&&\left.-\frac{(p-p')_\mu}{q^{\prime 2}-M_\gamma^2}T^{\lambda\mu}_{K\pi}(q';p',p)+\frac{i}{q^{\prime 2}-M_\gamma^2}
\Gamma^\lambda_{K\pi}(q';p',p)\right\}\nonumber\\
&\equiv&\frac{\alpha}{2\pi}\left[\ln\frac{M_W^2}{m_e^2}+\frac{3}{4}+\frac{1}{2}
\tilde{a}_g^\mathrm{res}\right]M_0+\left(\delta M_2+\delta M_{\gamma W}^a\right)_\mathrm{int}~.\label{eq:M2andMgammaWV}
\end{eqnarray}
The terms in the square bracket in Eq.\eqref{eq:M2andMgammaWV} are exactly known as they are isolated 
from the full one-loop integral with the help of the Ward identities in Eq.\eqref{eq:Ward}, as well
as the operator product expansion (OPE) at leading-twist in the $q^\prime\sim M_W$ region (see Eq.(6.1)
in Ref.\cite{Seng:2019lxf}), and $\tilde{a}_g^\mathrm{res}\approx 0.019$ entails the
$\mathcal{O}(\alpha_s)$ pQCD corrections of such terms. The remaining ``integral'' piece requires
further theoretical analysis and will be treated in the next section. Meanwhile, the other component
of the $\gamma W$-box diagram reads:
\begin{eqnarray}
\delta M_{\gamma W}^b&=&-i\frac{G_Fe^2}{\sqrt{2}}\bar{u}_\nu\gamma_\lambda(1-\gamma_5)v_e\int
\frac{d^4q'}{(2\pi)^4}\frac{M_W^2}{M_W^2-q^{\prime 2}}\frac{1}{(p_e-q')^2-m_e^2}\frac{1}{q^{\prime 2}}
\epsilon^{\mu\nu\alpha\lambda}q'_\alpha T_{\mu\nu}^{K\pi}(q';p',p)~,\nonumber\\
\label{eq:deltaMWb}
\end{eqnarray}
which can be split into two pieces, as well: $\delta M_{\gamma W}^b=\delta M_{\gamma W}^{b,V}+
\delta M_{\gamma W}^{b,A}$, where $\delta M_{\gamma W}^{b,V}$ ($\delta M_{\gamma W}^{b,A}$) picks up the
contribution from the vector (axial) charged weak current in the generalized Compton tensor
$T_{\mu\nu}^{K\pi}$.

At this point, we can combine the terms in the square brackets from Eqs.\eqref{eq:deltaMEW} and
\eqref{eq:M2andMgammaWV}. They are analytically known and do not require any further
treatment. Their contribution to $\delta f_+^{K\pi}$ is given by
\begin{equation}
\left(\delta f_+^{K\pi}\right)_\mathrm{I}=\left\{\frac{\alpha}{2\pi}\left[\ln\frac{M_Z^2}{m_e^2}
-\frac{1}{4}\ln\frac{M_W^2}{m_e^2}+\frac{1}{2}\ln\frac{m_e^2}{M_\gamma^2}-\frac{3}{8}
+\frac{1}{2}\tilde{a}_g\right]+\frac{1}{2}\delta_\mathrm{HO}^\mathrm{QED}\right\}f_+^{K\pi}(t)~,
\label{eq:deltafana}
\end{equation}
where $\tilde{a}_g=-(3/2)a_\mathrm{pQCD}+\tilde{a}_g^\mathrm{res}\approx -0.083$. We use the subscript
``I'' to signify the fact that it carries an IR~divergence. We will see later that the remaining
IR-divergent pieces in the virtual corrections come from $\left(\delta M_2+\delta M_{\gamma W}
\right)_\mathrm{int}$ and $\delta M_3$, and will carry the subscript ``II'' and ``III'', respectively.

All the remaining $\mathcal{O}(G_F\alpha)$ electroweak RC to the $n=0$ decay amplitude not included
in Eq.~\eqref{eq:deltafana} are fully contained in the following quantities: $\left(\delta M_2
+\delta M_{\gamma W}^a\right)_\mathrm{int}$, $\delta M_{\gamma W}^{b,V}$, $\delta M_3$ and
$\delta M_{\gamma W}^{b,A}$. They will be studied in the next three sections. 

\section{\label{sec:Born}
  Virtual correction: $\left(\delta M_2+\delta M_{\gamma W}^a\right)_\mathrm{int}$
  and $\delta M_{\gamma W}^{b,V}$}

\begin{figure}
	\begin{centering}
		\includegraphics[scale=0.3]{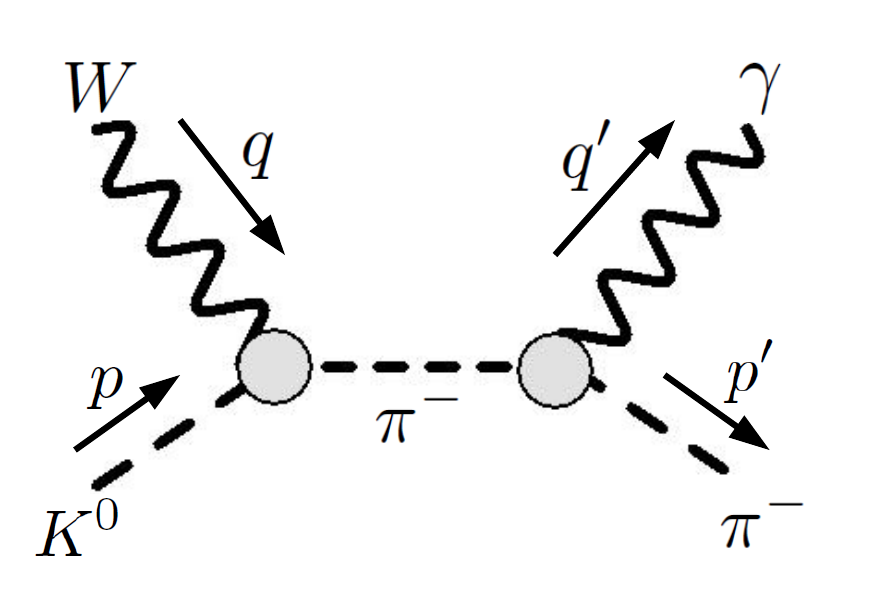}
		\includegraphics[scale=0.3]{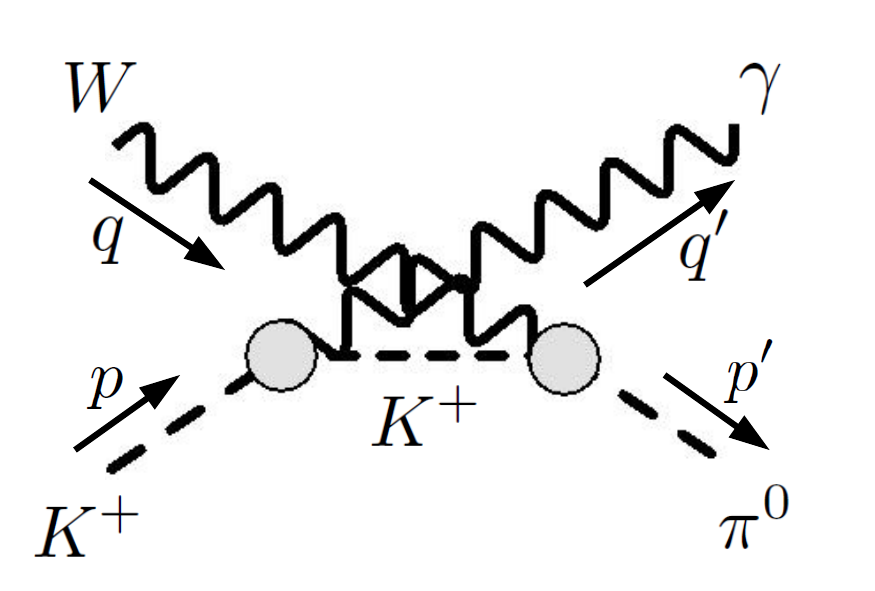}
		\includegraphics[scale=0.3]{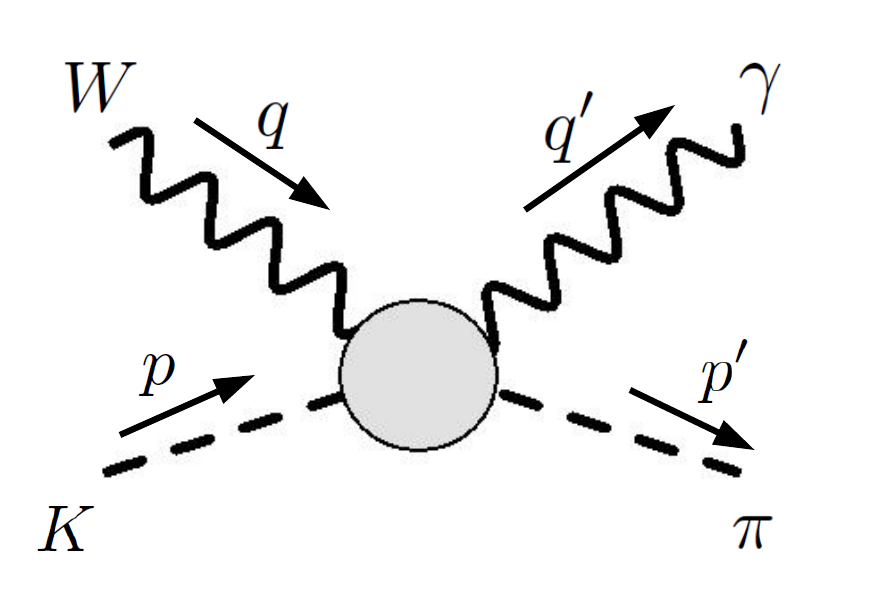}\hfill
		\par\end{centering}
	        \caption{\label{fig:TmunulowQ}Pole (left, middle) and seagull (right) contribution to
                  $T_{\mu\nu}^{K\pi}$ at low energy.}
\end{figure}

In this section we evaluate the loop integrals in
$\left(\delta M_2+\delta M_{\gamma W}^a\right)_\mathrm{int}$ and $\delta M_{\gamma W}^{b,V}$. The first
important observation is that these integrals cannot depend on physics at large virtual
momentum $q'$ (so we could take $M_W^2/(M_W^2-q^{\prime 2})\to 1$ in the integrand). In $\left(\delta M_2
+\delta M_{\gamma W}^a\right)_\mathrm{int}$, this is because the numerators in the integrand contain
explicit factors of $p_e$, $p-p'$ or quark masses (in $\Gamma^\lambda_{K\pi}$); whereas in
$\delta M_{\gamma W}^{b,V}$, it is because there is no extra antisymmetric tensor coming from
$\left(T_{\mu\nu}^{K\pi}\right)_V$, so the integral vanishes when $q'\gg (p-p')$ or $p_e$ due
to symmetry. Therefore, these integrals are saturated by contributions from the intermediate
hadronic states at low energy.

All the information on the hadronic structure in these integrals is contained in the generalized
Compton tensor $T^{\mu\nu}_{K\pi}$ and the vector $\Gamma_{K\pi}^\mu$. 
Within the former, we distinguish two types of contributions shown in Fig.\ref{fig:TmunulowQ}: the pole term associated with a charged meson 
(initial or final, depending on the reaction channel) propagator which leads to a $1/q'$ behavior in the soft photon limit, 
and the seagull term which is regular in that limit. The pole term is model-independent and given in terms of the meson 
weak and electromagnetic form factors, whereas the seagull term, alongside the form factors, contains information about excited states, 
and is generally model-dependent. It is common to single out the Born part of the generalized Compton tensor, defined as the pole terms 
complemented by a part of the seagull term that ensures that the Ward identities in Eq.\eqref{eq:Ward} are satisfied. In this way, the remaining, 
non-Born part is regular for $q'\to0$ and also obeys Ward identities individually.

Guided by the order $\mathcal{O}(p^2)$ result  in chiral expansion for the Compton tensor,
\begin{eqnarray}
\left(T^{\mu\nu}_{K^0\pi^-}\right)_{p^2}&=&iV_{us}^*\left[\frac{(2p'+q')^\mu(p+p'+q')^\nu}{(p'+q')^2-M_\pi^2}-g^{\mu\nu}\right]\nonumber\\
\left(T^{\mu\nu}_{K^+\pi^0}\right)_{p^2}&=&-\frac{iV_{us}^*}{\sqrt{2}}\left[\frac{(2p-q')^\mu(p+p'-q')^\nu}{(p-q')^2-M_K^2}-g^{\mu\nu}\right]~,\label{eq:TmunuLO}
\end{eqnarray}
and 
\begin{eqnarray}
\left(\Gamma_{K^0\pi^-}^\mu\right)_{p^2}&=&V_{us}^*\frac{M_K^2-M_\pi^2}{(p'+q')^2-M_\pi^2}(2p'+q')^\mu\nonumber\\
\left(\Gamma_{K^+\pi^0}^\mu\right)_{p^2}&=&-\frac{V_{us}^*}{\sqrt{2}}\frac{M_K^2-M_\pi^2}{(p-q')^2-M_K^2}(2p-q')^\mu~,\label{eq:GammaLO}
\end{eqnarray}
we thus define the {\color{black}minimal} Born contributions for the two decay channels as
\begin{eqnarray}
T^{\mu\nu,\,\rm B}_{K^0\pi^-}&=&iV_{us}^*F_\mathrm{em}^{\pi^-}(q^{\prime 2})
\left[\frac{(2p'+q')^\mu}{(p'+q')^2-M_\pi^2}
\left(f_+^{K^0\pi^-}(q^2)(2p+q)^\nu-f_-^{K^0\pi^-}(q^2)q^\nu\right)-g^{\mu\nu}f_{\rm seagull}^{K^0\pi^-}\right]
\nonumber\\
T^{\mu\nu,\,\rm B}_{K^+\pi^0}&=&iV_{us}^*F_\mathrm{em}^{K^+}(q^{\prime 2})
\left[\frac{(2p-q')^\mu}{(p-q')^2-M_K^2}
\left(f_+^{K^+\pi^0}(q^2)(2p'-q)^\nu-f_-^{K^+\pi^0}(q^2)q^\nu\right)-g^{\mu\nu}f_{\rm seagull}^{K^+\pi^0}
\right],
\nonumber\\\label{eq:TmunuBorn}
\end{eqnarray}
where $F_\mathrm{em}^{\pi^-}(q^{\prime 2})$ and $F_\mathrm{em}^{K^+}(q^{\prime 2})$ are the electromagnetic form
factors of the $\pi^-$ and the $K^+$, respectively\footnote{In principle the photon can also couple to $K^0$ due to
its non-zero charge radius, so $F_\mathrm{em}^{K^0}(q^{\prime 2})\neq 0$ when $q^{\prime 2}\neq 0$.
However, a simple ChPT calculation at $\mathcal{O}(p^4)$ indicates that
$|F_\mathrm{em}^{K^0}(q^{\prime 2})|<0.02$ when $|q^{\prime 2}|<0.1$~GeV$^2$ (see, e.g., Ref.\cite{Shi:2020rkz}),
so to our required precision it is completely negligible. On the other hand, $F_\mathrm{em}^{\pi^0}
(q^{\prime 2})$ is exactly zero due to $G$-parity.}, which satisfy $F_\mathrm{em}^{\pi^-}(0)=-1$ and
$F_\mathrm{em}^{K^+}(0)=1$. Furthermore, the normalization of the seagull term is fixed as:
\begin{eqnarray}
f_{\rm seagull}^{K^0\pi^-}=f_+^{K^0\pi^-}(q^2)-f_-^{K^0\pi^-}(q^2)\,,\quad\quad f_{\rm seagull}^{K^+\pi^0}=f_+^{K^+\pi^0}(q^2)+f_-^{K^+\pi^0}(q^2)\,.
\label{eq:SeagullFF}
\end{eqnarray}
One can check that the electromagnetic Ward identity is satisfied upon neglecting the $q'$-dependence of the form factors in $q_\mu' T^{\mu\nu,\mathrm{B}}_{K\pi}$.
With the same diagrams and keeping in mind that we must apply the
equation of motion to the charged weak vertex so that it vanishes exactly when $M_K=M_\pi$ (see, e.g.
the discussion in Sec.~7 of Ref.\cite{Seng:2019lxf}), the Born contribution to $\Gamma^\mu_{K\pi}$ reads, 
\begin{eqnarray}
\Gamma^{\mu,\,\rm B}_{K^0\pi^-}&=&V_{us}^*\frac{M_K^2-M_\pi^2}{(p'+q')^2-M_\pi^2}(2p'+q')^\mu
F_\mathrm{em}^{\pi^-}(q^{\prime 2})f_0^{K^0\pi^-}(q^2)\nonumber\\
\Gamma^{\mu,\,\rm B}_{K^+\pi^0}&=&V_{us}^*\frac{M_K^2-M_\pi^2}{(p-q')^2-M_K^2}(2p-q')^\mu
F_\mathrm{em}^{K^+}(q^{\prime 2})f_0^{K^+\pi^0}(q^2)\label{eq:GammaBorn}
\end{eqnarray} 
that depends on the scalar but not the vector charged weak form factor. 

{\color{black}The Born contributions of Eqs.\eqref{eq:TmunuBorn} are defined in terms of the model-independent pole contributions supplemented with a minimal seagull term required by gauge invariance. 
It is easy to see that if rearranging Eqs.\eqref{eq:TmunuBorn} into two separately gauge invariant structures (clearly reminiscent of the usual inelastic structure functions, $(-g^{\mu\nu}+\dots)F_1+(p^\mu p^\nu+\dots)/(p\cdot q)F_2$), one finds that only the contribution to $F_2$ contains a pole and is model-independent. The Born contribution to $F_1$ is regular and cannot in principle be distinguished from other inelastic contributions, so that Eqs.\eqref{eq:TmunuBorn} represent the minimal Born contribution definition only, bearing residual model dependence. Fortunately, its effect on the loop integrals turns out to be very small. 
In $\left(\delta M_2+\delta M_{\gamma W}^a\right)_\mathrm{int}$, it only contributes to $\delta f_-^{K\pi}$, whose effect in the decay rate is further suppressed by $r_e \approx10^{-6}$ (which is yet another reason why we restrict ourselves to $K_{e3}$ throughout this study), whereas the contribution to $\delta M_{\gamma W}^{b,V}$ vanishes trivially due to symmetry. 
	
Starting from $\mathcal{O}(p^4)$ one expects new structures such as $p^\mu p^\nu/\Lambda^2$ to enter, which parametrize inelastic contributions. Observe that a new mass scale $\Lambda$ is present for dimensional reasoning, and an obvious choice is the mass of the lowest resonances. This means we are able to get a handle of the effect of the inelastic contributions by computing the contributions from the resonances at low energy. We perform that
calculation based on the framework of resonance chiral theory (this is fine, as we are only dealing with tree graphs, see details in Appendix~\ref{sec:res}), and find that their contribution to $\delta_{K_{e3}}$ through $\left(\delta M_2+\delta M_{\gamma W}^a\right)_\mathrm{int}$ and
$\delta M_{\gamma W}^{b,V}$  is smaller than $10^{-4}$, which indicates that this  contribution is negligible. 
However, to stay on the safe side, we introduce a common uncertainty of $2\times 10^{-4}$, which is roughly four times the magnitude of the resonance contribution estimated in Appendix~\ref{sec:res}, to $\delta_{K_{e3}}$ as a very conservative estimation of the effects from the neglected inelastic terms. }

Before proceeding directly with the numerical calculations, we prefer to further isolate a
particularly important piece from $T_{K\pi}^{\mu\nu,\,\rm B}$ and
$\Gamma_{K\pi}^{\mu,\,\rm B}$ known as the ``convection term''~\cite{Meister:1963zz},
\begin{eqnarray}
T^{\mu\nu,\,\rm conv}_{K^0\pi^-}&=&
-\frac{i(2p'+q')^\mu F_{K^0\pi^-}^\nu(p',p)}{(p'+q')^2-M_\pi^2}\nonumber\\
T^{\mu\nu,\,\rm conv}_{K^+\pi^0}&=&\frac{i(2p-q')^\mu F_{K^+\pi^0}^\nu(p',p)}{(p-q')^2-M_K^2}~,
\label{eq:Tmunuconv}\\
\Gamma^{\mu,\,\rm conv}_{K^0\pi^-}&=&\frac{(2p'+q')^\mu (p'-p)_\lambda
F_{K^0\pi^-}^\lambda(p',p)}{(p'+q')^2-M_\pi^2}\nonumber\\
\Gamma^{\mu,\,\rm conv}_{K^+\pi^0}&=&-\frac{(2p-q')^\mu (p'-p)_\lambda
F_{K^+\pi^0}^\lambda(p',p)}{(p-q')^2-M_K^2}~.
\end{eqnarray}
It corresponds to taking the contribution of the point electric charge in the Born term. 
This contribution contains the full IR-divergent structure and is numerically the
largest. Being $q'$-independent, it leads to a contribution to the loop integrals that does not 
depend on the specific parameterization of the hadronic form factors. 
Therefore it gives rise to the so-called ``model-independent''
contribution emphasized in Refs.\cite{Garcia:1981it,JuarezLeon:2010tj,Torres:2012ge,Neri:2015eba},
which is more commonly known as the ``outer correction'' in the case of free neutron and nuclear
beta decays~\cite{Sirlin:1967zza,Wilkinson:1970cdv}. We thus choose to split the full Born
contribution to $\delta f_+^{K\pi}$ into three pieces as follows: 
\begin{equation}
\left(\delta f_+^{K\pi}\right)_\mathrm{Born}=\left(\delta f_+^{K\pi}\right)_\mathrm{II}
+\left(\delta f_+^{K\pi}\right)_\mathrm{conv}^\mathrm{fin}+\left(\delta f_+^{K\pi}\right)_\mathrm{Born-conv}~.
\label{eq:Bornsplit}
\end{equation}
The first and the second piece on the right-hand side of the equation above represent the IR-divergent
and IR-finite contributions from the convection term, respectively. The last piece,
$\left(\delta f_+^{K\pi}\right)_{\mathrm{Born-conv}}$, represents the difference between the full
Born contribution and the convection term contribution. In what follows we provide the analytic
results for the first two pieces: 
\begin{eqnarray}
\left(\delta f_+^{K^0\pi^-}\right)_\mathrm{II}&=&-\frac{\alpha}{4\pi}\left\{-\frac{4p_e\cdot p' x_s}{m_e M_\pi(1-x_s^2)}\ln x_s\ln\left(\frac{M_\gamma^2}{m_eM_\pi}\right)f_+^{K^0\pi^-}(t)\right.\nonumber\\
&&\left.+\left(\frac{5}{2}-\ln\frac{M_\pi^2}{M_\gamma^2}\right)\left(\frac{p'\cdot (p+p')}{2M_\pi^2}f_+^{K^0\pi^-}(t)+\frac{p'\cdot(p-p')}{2M_\pi^2}f_-^{K^0\pi^-}(t)\right)\right\}\nonumber\\
\left(\delta f_+^{K^+\pi^0}\right)_\mathrm{II}&=&-\frac{\alpha}{4\pi}\left\{\frac{4p_e\cdot p x_u}{m_e M_K(1-x_u^2)}\ln x_u\ln\left(\frac{M_\gamma^2}{m_eM_K}\right)f_+^{K^+\pi^0}(t)\right.\nonumber\\
&&\left.+\left(\frac{5}{2}-\ln\frac{M_K^2}{M_\gamma^2}\right)\left(\frac{p\cdot (p+p')}{2M_K^2}f_+^{K^+\pi^0}(t)+\frac{p\cdot(p-p')}{2M_K^2}f_-^{K^+\pi^0}(t)\right)\right\}~,\nonumber\\
\end{eqnarray}
and
\begin{eqnarray}
(\delta f_+^{K^0\pi^-})_\mathrm{conv}^\mathrm{fin}&=&-\frac{\alpha}{4\pi}\biggl\{\left(C_{00}^\mathrm{fin}+4p_e\cdot p'C_0^\mathrm{fin}+2p_e\cdot p' C_1-2m_e^2 C_2\right)f_+^{K^0\pi^-}(t)\biggr.\nonumber\\
&&+\left(p'\cdot(p+p')f_+^{K^0\pi^-}(t)+p'\cdot(p-p')f_-^{K^0\pi^-}(t)\right)\left(C_1+\frac{1}{2}C_{11}\right)\nonumber\\
&&-\frac{1}{2}\left(p_e\cdot(p+p')f_+^{K^0\pi^-}(t)+p_e\cdot(p-p')f_-^{K^0\pi^-}(t)\right)C_{12}\nonumber\\
&&\biggl.+\left(p_e\cdot(p'-p)+m_e^2\right)\left(f_+^{K^0\pi^-}(t)+f_-^{K^0\pi^-}(t)\right)C_2\biggr\}\nonumber\\
(\delta f_+^{K^+\pi^0})_\mathrm{conv}^\mathrm{fin}&=&-\frac{\alpha}{4\pi}\biggl\{\left(C_{00}^\mathrm{fin}-4p_e\cdot pC_0^\mathrm{fin}-2p_e\cdot p C_1-2m_e^2 C_2\right)f_+^{K^+\pi^0}(t)\biggr.\nonumber\\
&&+\left(p\cdot(p+p')f_+^{K^+\pi^0}+p\cdot(p-p')f_-^{K^+\pi^0}\right)\left(C_1+\frac{1}{2}C_{11}\right)\nonumber\\
&&+\frac{1}{2}\left(p_e\cdot(p+p')f_+^{K^+\pi^0}(t)+p_e\cdot(p-p')f_-^{K^+\pi^0}(t)\right)C_{12}\nonumber\\
&&\biggl.+\left(p_e\cdot(p'-p)+m_e^2\right)\left(f_+^{K^+\pi^0}(t)-f_-^{K^+\pi^0}(t)\right)C_2\biggr\}~.
\end{eqnarray}
The variables $x_s$, $x_u$ and the loop functions are defined in Appendix~\ref{sec:loop}. Notice that
one needs to substitute  $m_1=M_\pi$, $m_2=m_e$, and $v=s=(p'+p_e)^2$ in the $C$-functions for
the case of $K_{e3}^0$, and $m_1=M_K$, $m_2=m_e$, $v=u=(p-p_e)^2$ for the case of $K_{e3}^+$. 

Next, we shall study $\left(\delta f_+^{K\pi}\right)_\mathrm{Born-conv}$, which is the only piece that
requires a specific parameterization of the hadronic form factors in order to perform the loop
integral. Our first observation is that the Born contribution to $\delta f_+^{K\pi}$ is UV-finite
even without the form factors (it is UV-divergent for $\delta f_-^{K\pi}$ without the form factors,
which is however irrelevant for $K_{e3}$). Therefore, we expect the effect of the form factors to
receive a regular power suppression instead of a logarithmic enhancement. 

There are different ways of parameterizing the form factors which are practically indistinguishable
in the region $q'\sim p_e\sim p-p'\sim M_K-M_\pi$ relevant to the integrals. However, in practice a
simpler parameterization allows for a more straightforward evaluation of the loop integrals.
Therefore, in this work, we shall adopt the monopole representation for both the electromagnetic
and charged weak form factors. It is advantageous because the monopole resembles an ordinary
propagator, so the $q'$-integral reduces to standard Passarino-Veltman loop functions which can
be integrated numerically with respect to $\{y,z\}$.\footnote{Throughout this research we make
extensive use of \textit{Package-X}~\cite{Patel:2015tea,Patel:2016fam}. It is a \textit{Mathematica}
package that provides very efficiently all the analytic expressions of one-loop integrals that can
be directly applied to the numerical phase-space integration.} For the electromagnetic form factors,
we have:
\begin{equation}
F_\mathrm{em}^{\pi^-}(q^{\prime 2})=\frac{-1}{1-\frac{1}{6}\left\langle R_\pi^2\right\rangle q^{\prime 2}}~,
\:\:F_\mathrm{em}^{K^+}(q^{\prime 2})=\frac{1}{1-\frac{1}{6}\left\langle R_K^2\right\rangle q^{\prime 2}}~,
\end{equation}
where $\left\langle R_\pi^2\right\rangle$ and $\left\langle R_K^2\right\rangle$ are the mean-square
charge radius of $\pi^-$ and $K^+$, respectively\footnote{A general monopole form factor would read
$F = 1/(1-q^2/\Lambda^2)$. Here, we simply express the cut-off $\Lambda$ in terms of the charge
  radius, as we are interested in a precise low-energy representation.}. For the former, we use the result in Ref.\cite{Amendolia:1986wj}:
\begin{equation}
\left\langle R_\pi^2\right\rangle=(0.431\pm 0.010)~\mathrm{fm}^2
\end{equation}
because it was obtained through an experimental fit to the monopole form factor, which is what we adopt in this work. This value is consistent with the more recent determinations~\cite{Ananthanarayan:2017efc,Colangelo:2018mtw} as well as the PDG average~\cite{Zyla:2020zbs}, and the 2\% experimental uncertainty is completely
negligible in our analysis. The kaon mean-square charge radius, on the other hand,
was measured with a 15\% uncertainty~\cite{Amendolia:1986ui}:
\begin{equation}
\left\langle R_K^2\right\rangle=(0.34\pm 0.05)~\mathrm{fm}^2~,\label{eq:RK2}
\end{equation}
{\color{black}which agrees with monopole-SU(3) estimates (see, e.g., Ref.\cite{Ecker:1988te}).}
We will include this uncertainty later in our error analysis. Finally, for the vector and
scalar charged weak form factor, the monopole parameterization reads:
\begin{equation}
\bar{f}_+(q^2)=\frac{M_V^2}{M_V^2-q^2}~,\:\:\bar{f}_0(q^2)=\frac{M_S^2}{M_S^2-q^2}~,
\end{equation}
where the fitted vector and scalar pole masses are~\cite{Lazzeroni:2018glh}: 
\begin{equation}
M_V=(884.4\pm 7.4)~\mathrm{MeV}~,\:\:M_S=(1208.3\pm 52.1)~\mathrm{MeV}~.
\end{equation}
The uncertainties are less than 5\% and can be safely neglected in our analysis.

To end this section, we summarize in Table~\ref{tab:Born} the numerical contributions to $\delta_{K_{e3}}$ from the
different pieces in Eq.~\eqref{eq:Bornsplit} (except $\left(\delta f_+^{K\pi}\right)_\mathrm{II}$ that
we need to combine with other terms to achieve IR-finiteness). For the error analysis, we retain
only the uncertainties of the order $10^{-4}$ or larger which, in this case, only arise from
$\left\langle R_K^2\right\rangle$.  The first column represents the physical results, but we
also consider two other cases for comparison. In the second column, we retain only the
$\mathcal{O}(e^2p^2)$ contributions, which corresponds to taking $\bar{f}_+=\bar{f}_0=1$ and
$F_\mathrm{em}^{\pi^-}(q^{\prime 2})=-1$, $F_\mathrm{em}^{K^+}(q^{\prime 2})=1$. Comparing to the numbers
in the first column, we find the inclusion of form factors has a larger impact on the $\delta_{K_{e3}^+}$
than on the $\delta_{K_{e3}^0}$ channel. In fact, the amount of shift in the former exceeds the
estimated $\mathcal{O}(e^2p^4)$ uncertainty of $0.19\%$ in the ChPT analysis~\cite{Cirigliano:2008wn}.
This is understandable because the effect of the form factors scales typically as $M_i^2/M^2$,
where $M^2$ is the typical mass scale in the monopole parameterization, and $M_i$ is the mass of
the charged meson. In $K_{e3}^+$ we have $M_i=M_K$ so the numerical impact is larger. Finally,
in the third column, we consider an unphysical case where $M_K=1.1M_\pi$. We observe in this
case that $\left(\delta_{K_{e3}}\right)_\mathrm{conv}^\mathrm{fin}\gg\left(\delta_{K_{e3}}
\right)_\mathrm{Born-conv}$, which proves our previous assertion that the contribution from
the convection term dominates when the initial and final hadronic states are nearly degenerate.

\begin{table}
	\begin{centering}
		\begin{tabular}{|c|c|c|c|}
			\hline 
			& Full & $\mathcal{O}(e^{2}p^{2})$&$M_K=1.1M_\pi$\tabularnewline
			\hline 
			\hline 
			$\left(\delta_{K_{e3}^{0}}\right)_{\mathrm{conv}}^{\mathrm{fin}}$ & $-5.0\times10^{-3}$ & $-5.3\times 10^{-3}$&$3.08\times 10^{-2}$\tabularnewline
			\hline 
			$\left(\delta_{K_{e3}^{0}}\right)_{\mathrm{Born-conv}}$ & $4.1\times10^{-3}$ & $3.6\times 10^{-3}$&$1\times 10^{-4}$\tabularnewline
			\hline 
			$\left(\delta_{K_{e3}^{+}}\right)_{\mathrm{conv}}^{\mathrm{fin}}$ & $9.6\times10^{-3}$ & $9.2\times 10^{-3}$& $9.9\times 10^{-3}$\tabularnewline
			\hline 
			$\left(\delta_{K_{e3}^{+}}\right)_{\mathrm{Born-conv}}$ & $1(1)_{\left\langle R_{K}^{2}\right\rangle }\times10^{-4}$ & $-1.8\times 10^{-3}$& $-1\times 10^{-4}$\tabularnewline
			\hline 
		\end{tabular}
		\par\end{centering}
	        \caption{\label{tab:Born}The IR-finite Born contribution to $\delta M_2+\delta
                  M_{\gamma W}^a+\delta M_{\gamma W}^{b,V}$.}
\end{table}

\section{\label{sec:3pt}Virtual correction: {\boldmath$\delta M_3$}}

Next, we study $\delta M_3$, namely the ``three-point function'' correction to the charged weak
form factors. It was suggested in Ref.\cite{Seng:2019lxf} to calculate such contributions in fixed-order ChPT,
and we obtain the following results at $\mathcal{O}(e^2p^2)$:
\begin{eqnarray}
\left(\delta f_{+,3}^{K^{+}\pi^{0}}\right)_{e^2p^2}&=&-\frac{\alpha}{4\sqrt{2}\pi}\frac{p\cdot(p-p')}{2M_{K}^{2}}\left[\ln\frac{M_{K}^{2}}{M_{\gamma}^{2}}-\frac{5}{2}\right]+\left(\delta f_{+,3}^{K^{+}\pi^{0}}\right)_{e^2p^2}^{\mathrm{fin}}\nonumber\\
\left(\delta f_{+,3}^{K^{0}\pi^{-}}\right)_{e^2p^2}&=&\frac{\alpha}{4\pi}\frac{p'\cdot(p-p')}{2M_{\pi}^{2}}\left[\ln\frac{M_{\pi}^{2}}{M_{\gamma}^{2}}-\frac{5}{2}\right]+\left(f_{+,3}^{K^{0}\pi^{-}}\right)_{e^2p^2}^{\mathrm{fin}}~,\label{eq:deltaf3e2p2}
\end{eqnarray}
where the IR-finite pieces read:
\begin{eqnarray}
\left(\delta f_{+,3}^{K^{+}\pi^{0}}\right)_{e^2p^2}^{\mathrm{fin}}&=&-\frac{8\pi Z\alpha}{\sqrt{2}}\left[\frac{1}{2}\bar{h}_{K^{+}\pi^{0}}(t)+\bar{h}_{K^{0}\pi^{-}}(t)+\frac{3}{2}\bar{h}_{K^{+}\eta}(t)\right]\nonumber\\
&&+\frac{Z\alpha}{2\sqrt{2}\pi}\frac{M_{K}^{2}}{M_{\eta}^{2}-M_{\pi}^{2}}\left[1+\ln\frac{M_{K}^{2}}{\mu^{2}}\right]-\frac{4\pi\alpha}{\sqrt{2}}\left[-2K_{3}^{r}+K_{4}^{r}+\frac{2}{3}K_{5}^{r}+\frac{2}{3}K_{6}^{r}\right]\nonumber\\
&&-\frac{8\pi\alpha}{\sqrt{2}}\frac{M_{\pi}^{2}}{M_{\eta}^{2}-M_{\pi}^{2}}\left[-2K_{3}^{r}+K_{4}^{r}+\frac{2}{3}K_{5}^{r}+\frac{2}{3}K_{6}^{r}-\frac{2}{3}K_{9}^{r}-\frac{2}{3}K_{10}^{r}\right]\nonumber\\
\left(f_{+,3}^{K^{0}\pi^{-}}\right)_{e^2p^2}^{\mathrm{fin}}&=&-8\pi Z\alpha\left[\frac{1}{2}\bar{h}_{K^{+}\pi^{0}}(t)+\bar{h}_{K^{0}\pi^{-}}(t)+\frac{3}{2}\bar{h}_{K^{+}\eta}(t)\right]~.\label{eq:3ptfinite}
\end{eqnarray}
The parameter $Z\approx 0.8$ represents the short-distance electromagnetic effects that causes
the $M_{\pi^\pm}-M_{\pi^0}$ mass splitting, while $\{K_i^r\}$ are the $\mathcal{O}(e^2p^2)$ LECs
in the chiral Lagrangian with dynamical photons~\cite{Urech:1994hd}. Finally, the loop
functions $\bar{h}_{PQ}(t)$ are defined in Appendix~A of Ref.\cite{Seng:2019lxf}.

The strategy above has a caveat, namely: There is an IR-divergent piece in $\left(\delta
f_{+,3}^{K\pi}\right)$ that is numerically large, so its associated $\mathcal{O}(e^2p^4)$ uncertainty
can also be significant. Fortunately, it is straightforward to resum the IR-divergent piece
to all orders in the chiral power counting by appropriately putting back the charged weak form
factors based on two simple criteria as follows:
\begin{enumerate}
\item The combination $\ln(M_i^2/M_\gamma^2)-5/2$ originates from the convection term
  contribution and should stay intact after the resummation. This is apparent by noticing that
  the same combination appears also in $\left(\delta f_+^{K\pi}\right)_\mathrm{II}$.
\item As we will show in Section~\ref{sec:brem}, the IR-divergent piece from the bremsstrahlung
  contribution takes the following form:
	\begin{equation}
	  \delta |M|_\mathrm{brem}^2=\frac{\alpha}{\pi}\left(\frac{1}{\beta_i(0)}\tanh^{-1}\beta_i(0)
          -1\right)\ln\left[\frac{M_K^2}{M_\gamma^2}\right]|M_0|^2(0,y,z)+...~,\label{eq:bremIR}
	\end{equation}
	(the definition of $\delta|M|^2_\mathrm{brem}$ is given in Eq.\eqref{eq:deltaM2brem}) where
        $\beta_i(0)$ is the speed of the positron in the rest frame of the charged meson (i.e. $\pi^-$
        in $K_{e3}^0$ and $K^+$ in $K_{e3}^+$). The $M_\gamma$-dependence above must be canceled exactly
        by the corresponding $M_\gamma$-dependence in $\left(\delta f_+^{K\pi}\right)_\mathrm{I}$,
        $\left(\delta f_+^{K\pi}\right)_\mathrm{II}$ and the IR-divergent piece in $\delta f_{+,3}^{K\pi}$.
\end{enumerate}

The arguments above lead straightforwardly to the following expression for $\delta f_{+,3}^{K\pi}$:
\begin{equation}
\delta f_{+,3}^{K\pi}=\left(\delta f_{+}^{K\pi}\right)_\mathrm{III}+\left\{\left(\delta f_{+,3}^{K\pi}\right)_{e^2p^2}^{\mathrm{fin}}+\mathcal{O}(e^2p^4)\right\}~,\label{eq:deltaf3full}
\end{equation}
where the fully-resummed IR-divergent terms read:
\begin{eqnarray}
\left(\delta f_{+}^{K^+\pi^0}\right)_\mathrm{III}&=&\frac{\alpha}{4\pi}\frac{p\cdot(p-p')}{2M_{K}^{2}}\left[f_+^{K^+\pi^0}(t)-f_-^{K^+\pi^0}(t)\right]\left[\ln\frac{M_{K}^{2}}{M_{\gamma}^{2}}-\frac{5}{2}\right]\nonumber\\
\left(\delta f_{+}^{K^{0}\pi^{-}}\right)_\mathrm{III}&=&-\frac{\alpha}{4\pi}\frac{p'\cdot(p-p')}{2M_{\pi}^{2}}\left[f_+^{K^0\pi^-}(t)+f_-^{K^0\pi^-}(t)\right]\left[\ln\frac{M_{\pi}^{2}}{M_{\gamma}^{2}}-\frac{5}{2}\right]~,
\end{eqnarray}
while the IR-finite terms stay unchanged as in Eq.~\eqref{eq:3ptfinite}.
A significant advantage of Eq.~\eqref{eq:deltaf3full} over the $\mathcal{O}(e^2p^2)$ expression in
Eq.~\eqref{eq:deltaf3e2p2} is that now only the $\{...\}$ term involves a chiral expansion and
must be associated with an $\mathcal{O}(e^2p^4)$ uncertainty. 

We end this section by summarizing the numerical contribution from $\left(\delta
f_{+,3}^{K\pi}\right)_{e^2p^2}^\mathrm{fin}$ to the decay rate in Table~\ref{tab:3pt}. The
$\mathcal{O}(e^2p^4)$ uncertainty is obtained by multiplying the central value by
$M_K^2/\Lambda_\chi^2$. The numerical values of the LECs $\{K_i^r\}$ at $\mu=M_\rho$ are
obtained from Refs.\cite{Moussallam:1997xx,DescotesGenon:2005pw} (also summarized in
Ref.\cite{Bijnens:2014lea}), and we assign a 100\% uncertainty to the sum of the LEC contributions.  

\begin{table}
	\begin{centering}
		\begin{tabular}{|c|c|}
			\hline 
			& $\left(\delta_{K_{e3}}\right)_{\mathrm{3}}^{\mathrm{fin}}$\tabularnewline
			\hline 
			\hline 
			$K_{e3}^{0}$ & $0.5(1)_{e^{2}p^{4}}\times10^{-3}$\tabularnewline
			\hline 
			$K_{e3}^{+}$ & $1.4(3)_{e^{2}p^{4}}(8)_{\mathrm{LEC}}\times10^{-3}$\tabularnewline
			\hline 
		\end{tabular}
		\par\end{centering}
	\caption{\label{tab:3pt}The IR-finite contribution from the three-point function.}
\end{table}

\section{\label{sec:axialbox}Virtual correction: $\delta M_{\gamma W}^{b,A}$}

The last piece of the virtual corrections to $f_+^{K\pi}(t)$ comes from $\delta M_{\gamma W}^{b,A}$,
which is fundamentally different from those we studied in Section~\ref{sec:Born} and \ref{sec:3pt}
in the sense that it probes the strong interaction physics in $T_{K\pi}^{\mu\nu}$ from $Q^2\equiv -q^{\prime 2}
=0$ all the way up to $Q^2\sim M_W^2$. At large $Q$, one could perform a leading-twist, free-field OPE
that gives us the large electroweak logarithm, but this treatment breaks down at small  $Q$. Also,
due to parity, there is no Born contribution in $\delta M_{\gamma W}^{b,A}$ that can be easily accounted
for as in the previous two sections. Instead, one needs to deal with contributions from inelastic
intermediate states residing at $Q\sim \Lambda_\chi$ that are governed by non-perturbative QCD. In
the language of ChPT, their corresponding uncertainties are buried in the poorly-constrained
LECs $X_1$ and $\bar{X}_6^\mathrm{phys}$~\cite{Cirigliano:2001mk,Cirigliano:2004pv,Cirigliano:2008wn,Seng:2020jtz}. 

As we mentioned in the Introduction, an important breakthrough happened in early 2020 as lattice QCD
started to pick up its role in this subject.  
A series of first-principles calculations were performed to study the so-called
``forward axial $\gamma W$-box'' defined as follows:
\begin{equation}
  \Box_{\gamma W}^{VA}(\phi_i,\phi_f,M)\equiv\frac{ie^2}{2M^2}\int\frac{d^4q'}{(2\pi)^4}
  \frac{M_W^2}{M_W^2-q^{\prime 2}}\frac{1}{(q^{\prime 2})^2}\epsilon^{\mu\nu\alpha\beta}q'_\alpha
  p_{\phi\beta}\frac{T_{\mu\nu}^{if}(q';p_\phi,p_\phi)}{F_+^{if}(0)}~,\label{eq:forwardbox}
\end{equation}
where $\phi_i$ and $\phi_f$ are two degenerate hadrons with mass $M$, and carry the same
external momentum $p_\phi$, and $F_+^{if}(0)$ is the form factor $f_+^{if}(0)$ multiplied by the appropriate CKM matrix element. The first calculation of $\Box_{\gamma W}^{VA}(\pi^+,\pi^0,M_\pi)$ in
Ref.~\cite{Feng:2020zdc} led to the reduction of the RC uncertainty in the pion semileptonic decay
by a factor of three. Shortly after that, a new calculation of $\Box_{\gamma W}^{VA}(K^0,\pi^-,M_\pi)$ in
the flavor SU(3) limit was performed~\cite{Ma:2021azh} following the suggestion in Ref.\cite{Seng:2020jtz}. These
two calculations together provided an improved determination of the LECs $X_1$ and $\bar{X}_6^\mathrm{phys}$ that agrees with the values quoted in the earlier ChPT papers~\cite{Cirigliano:2001mk,Cirigliano:2003yr,Cirigliano:2008wn} within error bars, which suggests that the error assignment in the latter is reasonable.
However, in the pure ChPT representation, the major source of theory uncertainty in the long-range
electromagnetic corrections to $K_{e3}$ comes from $\mathcal{O}(e^2p^4)$ instead of the LECs.
Therefore, the significance of the calculations above was not fully revealed within the traditional
framework. 

In this section, we will demonstrate how the above-mentioned lattice QCD results play a decisive
role within the new theory framework, namely to pin down $\delta M_{\gamma W}^{b,A}$. 
We start by splitting the forward axial $\gamma W$-box into two pieces:
\begin{equation}
\Box_{\gamma W}^{VA}(\phi_i,\phi_f,M)=\Box_{\gamma W}^{VA>}+\Box_{\gamma W}^{VA<}(\phi_i,\phi_f,M)
\end{equation}
which come from the integral in Eq.\eqref{eq:forwardbox} at $Q^2> Q_\mathrm{cut}^2$ and
$Q^2<Q_\mathrm{cut}^2$, respectively, where $Q_\mathrm{cut}$ is a scale above which the
leading-twist, free-field OPE is applicable. Throughout this work we choose $Q^2_\mathrm{cut}=2$~GeV$^2$,
in accordance with the original lattice QCD paper~\cite{Feng:2020zdc}\footnote{The validity of
this choice is justified by the observation that the difference between the pQCD corrections to
$\mathcal{O}(\alpha_s^3)$ and to $\mathcal{O}(\alpha_s^4)$ is negligible above
$2$~GeV$^2$~\cite{Seng:2020wjq}, which demonstrates the convergence of the perturbative series.}.
The first term, $\Box_{\gamma W}^{VA>}$, contains a large electroweak logarithm and is independent
of the external states $\{\phi_i,\phi_f\}$ as well as the mass $M$. It is given by: 
\begin{equation}
\Box_{\gamma W}^{VA>}=\frac{\alpha}{8\pi}\ln\frac{M_W^2}{Q_\mathrm{cut}^2} + \ldots \,,
\end{equation}
where ``$+ \ldots$'' denotes the pQCD corrections, which are at present calculated to
$\mathcal{O}(\alpha_s^4)$~\cite{Baikov:2010je}, leading to a very precise determination: 
$\Box_{\gamma W}^{VA>}=2.16\times 10^{-3}$. Meanwhile, $\Box_{\gamma W}^{VA<}(\phi_i,\phi_f,M)$
depends $\{\phi_i,\phi_f,M\}$ and probe the details of the strong interaction at $Q\sim \Lambda_\chi$.  

To proceed further, we perform the same splitting to the integral in $\delta M_{\gamma W}^{b,A}$:
\begin{equation}
ie^2\int\frac{d^4q'}{(2\pi)^4}\frac{M_W^2}{M_W^2-q^{\prime 2}}\frac{1}{(p_e-q')^2-m_e^2}
\frac{1}{q^{\prime 2}}\epsilon^{\mu\nu\alpha\lambda}q'_\alpha \left(T_{\mu\nu}^{K\pi}(q';p',p)\right)_{A}
=\left\{\int_{Q^{2}>Q_\mathrm{cut}^2}+\int_{Q^2<Q_\mathrm{cut}^2}\right\}\left(...\right),
\end{equation}
where $\left(T_{\mu\nu}^{K\pi}\right)_A$ represents the component in $T_{\mu\nu}^{K\pi}$ that involves
the axial charged weak current. The contributions from these two terms to $\delta f_+^{K\pi}$
are denoted as $\left(\delta f_+^{K\pi}\right)_{\gamma W}^{b,A>}$ and $\left(\delta f_+^{K\pi}
\right)_{\gamma W}^{b,A<}$, respectively, and will now be related to the different components of the
forward axial $\gamma W$-box. First, since at $Q^2>Q^2_\mathrm{cut}\gg |p_e|^2$ we can
set $p_e\to 0$ in the integrand, one can show using OPE that,
\begin{equation}
\left(\delta f_+^{K\pi}\right)_{\gamma W}^{b,A>}=\Box_{\gamma W}^{VA>}f_+^{K\pi}(t)~.\label{eq:deltaf>}
\end{equation}
Adding this piece to $(\delta f_+^{K\pi})_\mathrm{I}$ in Eq.\eqref{eq:deltafana} reproduces
the full electroweak logarithm in the total RC. 

Next, we can parameterize the integral at $Q^2<Q_\mathrm{cut}^2$ as:
\begin{eqnarray}
&&ie^2\int_{Q^2<Q_\mathrm{cut}^2}\frac{d^4q'}{(2\pi)^4}\frac{M_W^2}{M_W^2-q^{\prime 2}}
\frac{1}{(p_e-q')^2-m_e^2}\frac{1}{q^{\prime 2}}\epsilon^{\mu\nu\alpha\lambda}q'_\alpha
\left(T_{\mu\nu}^{K\pi}(q';p',p)\right)_{A}\nonumber\\
&\equiv&V_{us}^*\left[g_+(M_K^2,M_\pi^2,m_e^2,s,u)(p+p')^\lambda+g_-(M_K^2,M_\pi^2,m_e^2,s,u)(p-p')^\lambda
\right.\nonumber\\
&&\left.+g_e(M_K^2,M_\pi^2,m_e^2,s,u)p_e^\lambda\right]~,\label{eq:Box<}
\end{eqnarray}
so it is obvious that:
\begin{equation}
\left(\delta f_+^{K\pi}\right)_{\gamma W}^{b,A<}=g_+(M_K^2,M_\pi^2,m_e^2,s,u)~.\label{eq:deltaf<}
\end{equation}
To relate this quantity to the recent lattice QCD results, we set $p\to p'$ and $p_e\to 0$ on both
sides of Eq.\eqref{eq:Box<}. That gives\footnote{In the last line we made two implicit approximations:
(1) we do not distinguish the value of $f_+^{K\pi}(0)$ between the case of $M_K>M_\pi$ and $M_K=M_\pi$,
and (2) we add the $t$-dependence to the form factor. Both approximations only lead to changes of
a few percent in $f_+^{K\pi}$, which is completely negligible after multiplying with
$\Box_{\gamma W}^{VA<}(K,\pi,M_\pi)$.}:
\begin{eqnarray}
g_+(M_\pi^2,M_\pi^2,0,M_\pi^2,M_\pi^2)&=&\frac{ie^2}{2M_\pi^2}\int_{Q^2<Q_\mathrm{cut}^2}\frac{d^4q'}{(2\pi)^4}
\frac{M_W^2}{M_W^2-q^{\prime 2}}\frac{1}{(q^{\prime 2})^2}\epsilon^{\mu\nu\alpha\lambda}q'_\alpha p'_\lambda\frac{\left(T_{\mu\nu}^{K\pi}(q';p',p')\right)_{A}}{V_{us}^*}\nonumber\\
&=&\Box_{\gamma W}^{VA<}(K,\pi,M_\pi)f_+^{K\pi}(t).
\end{eqnarray}
Since the lattice community has computed $\Box_{\gamma W}^{VA<}(K,\pi,M_\pi)$, we can obtain
$g_+(M_\pi^2,M_\pi^2,0,M_\pi^2,M_\pi^2)$ which is not exactly the same as $g_+(M_K^2,M_\pi^2,m_e^2,s,u)$
that we seek. However, remember that the integral in Eq.~\eqref{eq:Box<} is dominated by the
physics at the scale $q'\sim \Lambda_\chi$ (e.g. Regge physics~\cite{Seng:2020wjq}), it is then
possible to simply take $g_+(M_\pi^2,M_\pi^2,0,M_\pi^2,M_\pi^2)$ together with an
appropriately-assigned uncertainty:
\begin{equation}
g_+(M_K^2,M_\pi^2,m_e^2,s,u)=g_+(M_\pi^2,M_\pi^2,0,M_\pi^2,M_\pi^2)+\mathcal{O}\left(\frac{E^2}{\Lambda_\chi^2}\right)~,\label{eq:gplus}
\end{equation}
where $E$ is an energy scale that characterizes the non-forward (NF) kinematics in Eq.\eqref{eq:Box<},
e.g. $M_K-M_\pi$, $\left(s-M_\pi\right)^{1/2}$ or $\left(u-M_\pi\right)^{1/2}$. Since they are all smaller
than $M_K$, we can take $E\to M_K$ as a conservative estimation of the uncertainty due to the NF effects.
So, combining Eqs.\eqref{eq:deltaf>},~\eqref{eq:deltaf<} and \eqref{eq:gplus}, we obtain:
\begin{equation}
\left(\delta f_+^{K\pi}\right)_{\gamma W}^{b,A}=\left\{\Box_{\gamma W}^{VA>}
+\left[\Box_{\gamma W}^{VA<}(K,\pi,M_\pi)+\mathcal{O}\left(\frac{M_K^2}{\Lambda_\chi^2}\right)\right]\right\}
f_+^{K\pi}(t)~.
\end{equation}
Notice that only the term in the square bracket is associated to an $\mathcal{O}(M_K^2/\Lambda_\chi^2)$
uncertainty. 

The recent lattice calculations provided the forward axial $\gamma W$-box in the charged pion
and neutral kaon decay:
\begin{equation}
\Box_{\gamma W}^{VA<}(\pi^+,\pi^0,M_\pi)=0.671(28)_\mathrm{lat}\times 10^{-3},\:\:\:
\Box_{\gamma W}^{VA<}(K^0,\pi^-,M_\pi)=0.278(44)_\mathrm{lat}\times 10^{-3}~.\label{eq:Boxlattice}
\end{equation}
The box diagram in charged kaon decay is not yet computed, but can be related to the first two
through a matching to the $\mathcal{O}(e^2p^2)$ ChPT expression:
\begin{equation}
\Box_{\gamma W}^{VA<}(K^+,\pi^0,M_\pi)=2\Box_{\gamma W}^{VA<}(\pi^+,\pi^0,M_\pi)-\Box_{\gamma W}^{VA<}(K^0,\pi^-,M_\pi)
= 1.064(71)_\mathrm{lat}\times 10^{-3}~.
\end{equation}
The higher-order ChPT corrections to the expression above scales as $\mathcal{O}(M_\pi^2/\Lambda_\chi^2)$
and can be safely neglected in our error analysis\footnote{Nevertheless, a direct lattice calculation
of $\Box_{\gamma W}^{VA}(K^+,\pi^0,M_\pi)$ in the future is still very much desirable as it provides
an excellent test of the convergence speed of the chiral expansion in the SU(3) limit.}.
With the numbers above, we obtain the numerical correction to the $K_{e3}$ decay rate from
$\delta M_{\gamma W}^{b,A}$, as summarized in Table~\ref{tab:axialgammaW}. Notice that the
NF uncertainty is obtained by simply multiplying $2\Box_{\gamma W}^{VA<}(K,\pi,M_\pi)$
with $M_K^2/\Lambda_\chi^2$. 

\begin{table}
	\begin{centering}
		\begin{tabular}{|c|c|c|c|}
			\hline 
			$\left(\delta_{K_{e3}}\right)_{\gamma W}^{b,A}$&$>$&$<$& Total\tabularnewline
			\hline 
			\hline 
			$K_{e3}^{0}$&$4.3\times 10^{-3}$&$0.6(1)_\mathrm{lat}(1)_\mathrm{NF}\times 10^{-3}$ & $4.9(1)_{\mathrm{lat}}(1)_{\mathrm{NF}}\times10^{-3}$\tabularnewline
			\hline 
			$K_{e3}^{+}$&$4.3\times 10^{-3}$&$2.1(1)_\mathrm{lat}(4)_\mathrm{NF}\times 10^{-3}$ & $6.4(1)_{\mathrm{lat}}(4)_{\mathrm{NF}}\times10^{-3}$\tabularnewline
			\hline 
		\end{tabular}
		\par\end{centering}
	\caption{\label{tab:axialgammaW}Contribution from $\delta M_{\gamma W}^{b,A}$.}
	
\end{table}

To end this section, we briefly discuss the future role of the lattice QCD. The estimation of the NF
uncertainty in Eq.\eqref{eq:gplus} is physically sound but can be further improved with an extra
lattice calculation. This can be seen by considering the following relations:
\begin{eqnarray}
-\frac{8}{3}X_1+\bar{X}_6^\mathrm{phys}(M_\rho)&=&-\frac{1}{2\pi\alpha}\left(\Box_{\gamma W}^{VA}(K^0,\pi^-,M_\pi)-\frac{\alpha}{8\pi}\ln\frac{M_W^2}{M_\rho^2}\right)+\frac{1}{8\pi^2}\left(\frac{5}{4}-\tilde{a}_g\right)+\mathcal{O}\left(\frac{M_\pi^2}{\Lambda_\chi^2}\right)\nonumber\\
-\frac{8}{3}X_1+\bar{X}_6^\mathrm{phys}(M_\rho)&=&-\frac{1}{2\pi\alpha}\left(\Box_{\gamma W}^{VA}(K^+,K^0,M_K)-\frac{\alpha}{8\pi}\ln\frac{M_W^2}{M_\rho^2}\right)+\frac{1}{8\pi^2}\left(\frac{5}{4}-\tilde{a}_g\right)+\mathcal{O}\left(\frac{M_K^2}{\Lambda_\chi^2}\right)~.\nonumber\\
\end{eqnarray}
Both equations are obtained through a matching between the calculation of the RC based on Sirlin's approach and ChPT; the first line was given in Ref.\cite{Seng:2020jtz}
and the second line can be derived accordingly. We see that both $\Box_{\gamma W}^{VA}(K^0,\pi^-,M_\pi)$
and $\Box_{\gamma W}^{VA}(K^+,K^0,M_k)$ are matched to the same combination of LECs, except that the
latter is subject to larger higher-order corrections because the involved meson mass is $M_K$ which
is larger. That means, the difference in the numerical values between $\Box_{\gamma W}^{VA}(K^0,\pi^-,M_\pi)$
and $\Box_{\gamma W}^{VA}(K^+,K^0,M_k)$ provides an estimation of the size of the NF corrections in
Eq.\eqref{eq:gplus}. This strategy is very similar to the standard lattice QCD technique to estimate
the size of the chiral power corrections through the variation of the quark masses.  

\section{\label{sec:brem}Bremsstrahlung contribution}

\begin{figure}
	\begin{centering}
		\includegraphics[scale=0.4]{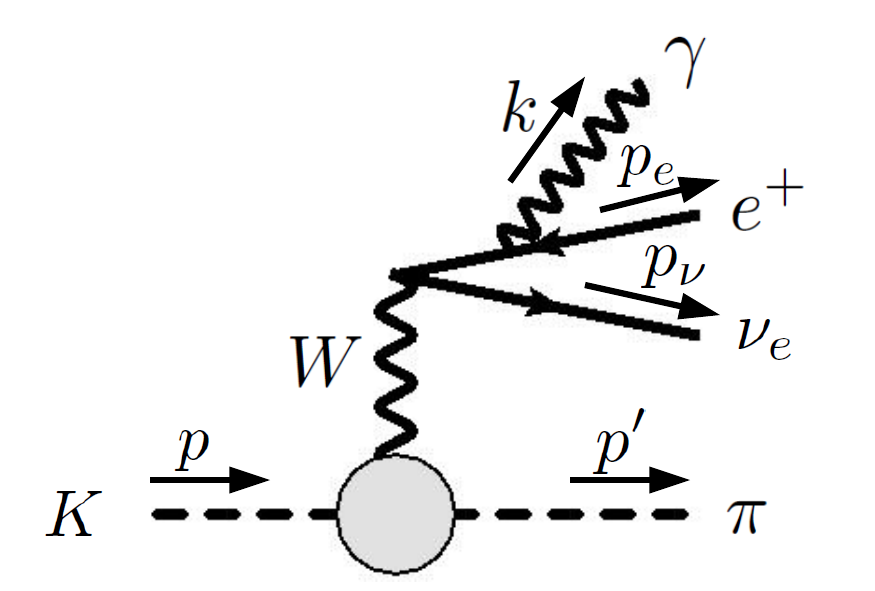}
		\includegraphics[scale=0.4]{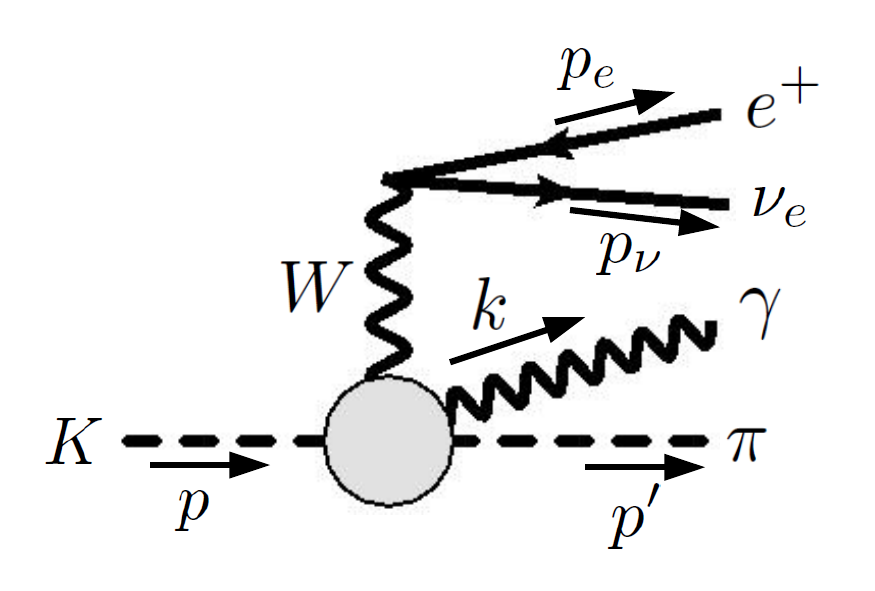}\hfill
		\par\end{centering}
	\caption{\label{fig:brems}The real photon emission diagrams.}
\end{figure}
 
After going through all the virtual corrections, we switch to the contribution from the $n=1$
process, which is simply known as the ``bremsstrahlung contribution''. According to the discussions in
Appendix~\ref{sec:PS}, the bremsstrahlung process contributes to the differential decay width
$d\Gamma_{K_{e3}}/dydz$ not only in the $\mathcal{D}_3$ region but also in the $\mathcal{D}_{4-3}$ region,
the latter has no correspondence in the $n=0$ process. Therefore, it is eventually up to the
experimentalists to decide in which region of $\{y,z\}$ will the data be taken, and whether or not a
veto will be applied to exclude decay events with hard photons. Of course, the simplest choice
is to not apply any veto, and to collect data from all available regions of $\{y,z\}$. This
corresponds to a fully-inclusive prescription of the real photon emission process, or in other
words, we should calculate the sum of the full $n=0$ and $n=1$ decay width. This
prescription was adopted in Ref.\cite{Cirigliano:2008wn} and will be followed in this work.

The bremsstrahlung amplitude, depicted by the two diagrams in Fig.\ref{fig:brems}, reads:
\begin{eqnarray}
M_{K\rightarrow\pi e^+\nu\gamma}&=&-\frac{G_Fe}{\sqrt{2}}\bar{u}_\nu\gamma^\mu(1-\gamma_5)\left
\{\frac{p_e\cdot\varepsilon^*(k)}{p_e\cdot k}+\frac{\slashed{k}\slashed{\varepsilon}^*(k)}{2p_e\cdot k}
\right\}v_e F_\mu^{K\pi}(p',p)\nonumber\\
&&+\frac{iG_Fe}{\sqrt{2}}\bar{u}_\nu\gamma^\nu(1-\gamma_5)v_e\varepsilon^{\mu *}(k)T_{\mu\nu}^{K\pi}(k;p',p)~.
\end{eqnarray}
We observe that the generalized Compton tensor $T_{\mu\nu}^{K\pi}$ appears again, only that now one
deals with a real photon. Unlike in the loop diagrams, here we only need to know $T_{\mu\nu}^{K\pi}$
for small (due to the phase-space constraint) and on-shell photon momentum $k$, so instead of
exhausting the contributions from all intermediate states, it is possible to adopt a
low-energy effective expression $T_{\mu\nu}^{K\pi}$. It should, however, satisfy three basic criteria:
\begin{itemize}
\item It must contain the full convection term contribution to ensure an exact cancellation
  of the IR-divergence from the virtual corrections.
\item It should include the seagull term, as the effect of the latter is not particularly suppressed
  in the decay rate, unlike in the loop diagrams.
\item It should satisfy exact electromagnetic gauge invariance, so that one could perform the usual
  replacement $\sum_s \varepsilon^\mu_s(k)\varepsilon^{\nu *}_s(k)\rightarrow -g^{\mu\nu}$ in the sum of
  the outgoing photon polarizations. 
\end{itemize}
The simplest effective expression that satisfies all these criteria is:
\begin{eqnarray}
T^{\mu\nu}_{K^0\pi^-}(k;p',p)&=&-\frac{i(2p'+k)^\mu F^\nu_{K^0\pi^-}(p',p)}{(p'+k)^2-M_\pi^2}
+\left\{iV_{us}^*\left[\frac{(2p'+k)^\mu k^\nu}{(p'+k)^2-M_\pi^2}-g^{\mu\nu}\right]+\mathcal{O}(p^4)\right\}
\nonumber\\
T^{\mu\nu}_{K^+\pi^0}(k;p',p)&=&\frac{i(2p-k)^\mu F^\nu_{K^+\pi^0}(p',p)}{(p-k)^2-M_K^2}+\left
\{\frac{iV_{us}^*}{\sqrt{2}}\left[\frac{(2p-k)^\mu k^\nu}{(p-k)^2-M_K^2}+g^{\mu\nu}\right]
+\mathcal{O}(p^4)\right\}.\label{eq:Tmunusplit}
\end{eqnarray}
The first term on the right-hand side in the expressions above is just the convection term, whereas
the remainders are the seagull term and the extra pieces from the Born contribution needed to
recover gauge invariance. Notice that the convection term is exact, and only the terms in
the curly bracket undergo a chiral expansion. In fact, if we expand the convection term to
$\mathcal{O}(p^2)$, the LO ChPT expression in Eq.\eqref{eq:TmunuLO} is recovered. In fact,
the existing ChPT calculation uses exactly Eq.\eqref{eq:TmunuLO} in their calculation of the
bremsstrahlung effect, but now our expression allows a resummation of the most important
terms in $T^{\mu\nu}_{K\pi}$ to all chiral orders.

With the above, the bremsstrahlung amplitude splits into two pieces: $M_{K\rightarrow\pi e^+\nu\gamma}=M_A+M_B$
that are separately gauge-invariant (i.e. we can write $M_{A,B}=\varepsilon_\mu^*(k)\tilde{M}_{A,B}^\mu$,
where $k_\mu \tilde{M}_{A,B}^\mu=0$). For $K^0\to \pi^- e^+\nu_e\gamma$ we have:
\begin{eqnarray}
M_A&=&-\frac{eG_F}{\sqrt{2}}F^{K^0\pi^-}_\mu(p',p)\varepsilon_\nu^*(k)\left\{\left(\frac{p_e}{p_e\cdot k}-\frac{p'}{p'\cdot k}\right)^\nu\bar{u}_\nu\gamma^\mu(1-\gamma_5)v_e+\frac{1}{2p_e\cdot k}\bar{u}_\nu\gamma^\mu(1-\gamma_5)\slashed{k}\gamma^\nu v_e\right\}\nonumber\\
M_B&=&-\frac{eG_F}{\sqrt{2}}V_{us}^*\varepsilon_\mu^*(k)\bar{u}_\nu\left\{\frac{p^{\prime\mu}}{p'\cdot k}\slashed{k}-\gamma^\mu\right\}(1-\gamma_5)v_e~,
\end{eqnarray}
and for $K^+\to \pi^0 e^+\nu_e\gamma$,
\begin{eqnarray}
M_A&=&-\frac{eG_F}{\sqrt{2}}F^{K^+\pi^0}_\mu(p',p)\varepsilon_\nu^*(k)\left\{\left(\frac{p_e}{p_e\cdot k}-\frac{p}{p\cdot k}\right)^\nu\bar{u}_\nu\gamma^\mu(1-\gamma_5)v_e+\frac{1}{2p_e\cdot k}\bar{u}_\nu\gamma^\mu(1-\gamma_5)\slashed{k}\gamma^\nu v_e\right\}\nonumber\\
M_B&=&\frac{eG_F}{2}V_{us}^*\varepsilon_\mu^*(k)\bar{u}_\nu\left\{\frac{p^{\mu}}{p\cdot k}\slashed{k}-\gamma^\mu\right\}(1-\gamma_5)v_e~.\label{eq:MAMBKp}
\end{eqnarray}
The significance of such a splitting is that $M_A$ is an exact expression and only $M_B$ involves a
chiral expansion. Therefore, in the computation of the decay rate, only the contribution from
$2\mathfrak{Re}\{M_B^* M_A\}+|M_B|^2$ acquires an $\mathcal{O}(e^2p^4)$ uncertainty, while the
contribution from $|M_A|^2$ is exact. As we will show later, this brings an advantage over the
existing treatment as the latter is numerically the largest. 

Now we proceed to the phase space integration of the bremsstrahlung contribution. We first discuss
the integration in the $\mathcal{D}_3$ region.
To isolate the IR-singular term, we first split $|M_A|^2$ into two pieces:
\begin{equation}
|M_A|^2=-e^2\left(\frac{p_e}{p_e\cdot k}-\frac{p_i}{p_i\cdot k}\right)^2|M_0|^2(0,y,z)+|M_A|^2_\mathrm{res},
\end{equation}
where $p_i=p$ ($p'$) in $K_{e3}^+$ ($K_{e3}^0$). The integration of the first term with respect
to $\{\vec{p}_\nu,\vec{k},x\}$ produces an IR-divergence:
\begin{equation}
\int_0^{\alpha_+(y,z)}dx\int\frac{d^3k}{(2\pi)^32E_k}\frac{d^3p_\nu}{(2\pi)^32E_\nu}(2\pi)^4\delta^{(4)}(P-k-p_\nu)\left(\frac{p_e}{p_e\cdot k}-\frac{p_i}{p_i\cdot k}\right)^2=I_i^\mathrm{IR}(y,z)+I_i^\mathrm{fin}(y,z)~,
\end{equation}
where the explicit expressions of $I_i^\mathrm{IR}$ and $I_i^\mathrm{fin}$ can be found in
Appendix~\ref{sec:DRIR}, and with this, we verify our previous assertion about the IR-divergent
structure of the bremsstrahlung contribution in Eq.\eqref{eq:bremIR}. We can now combine the IR-divergent
contributions from the virtual corrections (which we previously labeled as I, II, III) with the
bremsstrahlung contribution in the $\mathcal{D}_3$ region to obtain the following shift of the
$K_{e3}$ decay rate:
\begin{equation}
\left(\delta\Gamma_{K_{e3}}\right)_\mathrm{I,II,III+brem(\mathcal{D}_3)}=\frac{M_K}{256\pi^3}\int_{\mathrm{D}_3}dydz\delta|M|^2_\mathrm{I,II,III+brem(\mathcal{D}_3)}(y,z)~,
\end{equation}
where
\begin{eqnarray}
\delta |M|^2_\mathrm{I,II,III+brem(\mathcal{D}_3)}(y,z)&=&\left\{\frac{\alpha}{2\pi}\left[2\ln\frac{M_Z^2}{m_e^2}-\frac{1}{2}\ln\frac{M_W^2}{m_e^2}+\left(1-\frac{2}{\beta_i(0)}\tanh^{-1}\beta_i(0)\right)\ln\frac{M_i^2}{M_K^2}\right.\right.\nonumber\\
&&\left.\left.+\frac{1}{\beta_i(0)}\tanh^{-1}\beta_i(0)\ln\frac{M_i^2}{m_e^2}-\frac{13}{4}+\tilde{a}_g\right]+\delta_\mathrm{HO}^\mathrm{QED}-\frac{e^2M_K^2}{2\pi}I_i^\mathrm{fin}(y,z)\right\}\nonumber\\
&&\times|M_0|^2(0,y,z)+\frac{M_K^2}{2\pi}\int_0^{\alpha_+}dx\int\frac{d^3k}{(2\pi)^32E_k}\frac{d^3p_\nu}{(2\pi)^32E_\nu}\nonumber\\
&&\times(2\pi)^4\delta^{(4)}(P-k-p_\nu)\left\{|M_A|^2_\mathrm{res}+2\mathfrak{Re}\left\{M_A^*M_B\right\}+|M_B|^2\right\}~,
\end{eqnarray}
which is now explicitly IR-finite. We observe that the expression above still contains a residual
integral with respect to $\{\vec{p}_\nu,\vec{k},x\}$, but it is IR-finite and therefore can be
straightforwardly carried out with the method outlined in Appendix~\ref{sec:IRfin}.
The numerical result is summarized in Table~\ref{tab:bremD3}. The HO uncertainty comes from
$\delta_\mathrm{HO}^\mathrm{QED}$, while the $\mathcal{O}(e^2p^4)$ uncertainty is obtained by multiplying
the contribution from $2\mathfrak{Re}\left\{M_A^*M_B\right\}+|M_B|^2$ by $M_K^2/\Lambda_\chi^2$. We see
that these uncertainties are as small as $10^{-4}$, which is a clear success of our strategy
in the splitting of $T_{K\pi}^{\mu\nu}(k';p',p)$ in Eq.\eqref{eq:Tmunusplit}. 

\begin{table}
	\begin{centering}
		\begin{tabular}{|c|c|c|c|}
			\hline 
			$\left(\delta_{K_{e3}}\right)_{\mathrm{I,II,III+brem(\mathcal{D}_3)}}$&From $2\mathfrak{Re}\left\{M_A^*M_B\right\}+|M_B^2|$&Remainder& Total\tabularnewline
			\hline 
			\hline 
			$K_{e3}^{0}$&$0.10(2)_{e^2p^4}\times 10^{-2}$&$2.41(3)_\mathrm{HO}\times 10^{-2}$ & $2.51(3)_{\mathrm{HO}}(2)_{e^{2}p^{4}}\times10^{-2}$\tabularnewline
			\hline 
			$K_{e3}^{+}$&$-0.03(1)_{e^2p^4}\times 10^{-2}$&$0.44(3)_\mathrm{HO}\times 10^{-2}$ & $0.40(3)_{\mathrm{HO}}(1)_{e^{2}p^{4}}\times10^{-2}$\tabularnewline
			\hline 
		\end{tabular}
		\par\end{centering}
	\caption{\label{tab:bremD3}Sum of the IR-divergent one-loop contribution I, II, III
		and the bremsstrahlung contribution in the $\mathcal{D}_3$ region.}
\end{table}

Finally, we also need to compute the bremsstrahlung contribution in the $\mathcal{D}_{4-3}$ region:
\begin{eqnarray}
\left(\delta \Gamma_{K_{e3}}\right)_\mathrm{brem(\mathcal{D}_{4-3})}&=&\frac{M_K^3}{512\pi^4}\int_{\mathcal{D}_{4-3}}dydz\int_{\alpha_-(y,z)}^{\alpha_+(y,z)}dx\int\frac{d^3k}{(2\pi)^32E_k}\frac{d^3p_\nu}{(2\pi)^32E_\nu}(2\pi)^4\delta^{(4)}(P-k-p_\nu)\nonumber\\
&&\times\left\{|M_A|^2+2\mathfrak{Re}\left\{M_A^*M_B\right\}+|M_B|^2\right\}~.
\end{eqnarray}
The integrals are IR-finite and can be carried out similarly using the method in Appendix~\ref{sec:IRfin}.
The numerical results are given in Table~\ref{tab:bremD4m3}. In principle one also acquires an
$\mathcal{O}(e^2p^4)$ uncertainty by multiplying the contribution from $2\mathfrak{Re}\left
\{M_A^*M_B\right\}+|M_B|^2$ by $M_K^2/\Lambda_\chi^2$, but the outcomes are of the order $10^{-5}$
and so are not displayed in the table.

\begin{table}
	\begin{centering}
		\begin{tabular}{|c|c|c|c|}
			\hline 
			$\left(\delta_{K_{e3}}\right)_{\mathrm{brem(\mathcal{D}_{4-3})}}$&From $2\mathfrak{Re}\left\{M_A^*M_B\right\}+|M_B^2|$&From $|M_A|^2$& Total\tabularnewline
			\hline 
			\hline 
			$K_{e3}^{0}$&$0.2\times 10^{-3}$&$5.6\times 10^{-3}$ & $5.8\times10^{-3}$\tabularnewline
			\hline 
			$K_{e3}^{+}$&$-0.1\times 10^{-3}$&$5.3\times 10^{-3}$ & $5.2\times10^{-3}$\tabularnewline
			\hline 
		\end{tabular}
		\par\end{centering}
	\caption{\label{tab:bremD4m3}The bremsstrahlung contribution in the $\mathcal{D}_{4-3}$ region. Uncertainties are of order $10^{-5}$ and are not displayed.}
\end{table}

\section{\label{sec:comparison}Comparing with the ChPT result}
 
We have now finished calculating all components of the $\mathcal{O}(G_F^2\alpha)$ electroweak RC to
the $K_{e3}$ decay rate. The total result is simply given by:
\begin{equation}
\left(\delta_{K_{e3}}\right)_\mathrm{tot}=\left(\delta_{K_{e3}}\right)_\mathrm{conv}^\mathrm{fin}
+\left(\delta_{K_{e3}}\right)_\mathrm{Born-conv}+\left(\delta_{K_{e3}}\right)_3^\mathrm{fin}
+\left(\delta_{K_{e3}}\right)_{\gamma W}^{b,A}
+\left(\delta_{K_{e3}}\right)_{\mathrm{I,II,III+brem(\mathcal{D}_3)}}
+\left(\delta_{K_{e3}}\right)_\mathrm{brem(\mathcal{D}_{4-3})}~,
\end{equation}
where the numerical values of different components can be found in Tables~\ref{tab:Born}--\ref{tab:bremD4m3}. On the other hand, in the existing standard
ChPT treatment the full electroweak RC is broken down into
``short-distance'' and ``long-distance'' pieces, and are allocated to several different quantities,
some of which are somewhat implicitly hidden. This section serves to perform a rigorous matching between
our result and the values quoted in the existing ChPT literature, with special attention paid to
the so-called ``long-distance electromagnetic corrections'' $\delta_\mathrm{EM}^{Ke}$. 

In the standard ChPT framework, the photon-inclusive $K_{e3}$ decay rate is parameterized
as~\cite{Zyla:2020zbs}:
\begin{equation}
\Gamma_{K_{e3}}=\frac{G_F^2|V_{us}|^2M_K^5C_K^2}{192\pi^3}S_\mathrm{EW}|f_+^{K^0\pi^-}(0)|^2I_{Ke}^{(0)}(\lambda_i)
\left(1+\delta_\mathrm{EM}^{Ke}+\delta_\mathrm{SU(2)}^{K\pi}\right)~,
\end{equation}
where $C_K$ is a simple isospin factor. Apart from the quantity $|f_+^{K^0\pi^-}(0)|$ that requires a
lattice input, all the small QCD and electroweak corrections to $\Gamma_{K_{e3}}$ are distributed
into the following four quantities: $S_\mathrm{EW}$, $I_{Ke}^{(0)}(\lambda_i)$,  $\delta_\mathrm{SU(2)}^{K\pi}$
and $\delta_\mathrm{EM}^{Ke}$. We shall take a serious look at each of these quantities, and study
their relations to the different components of electroweak RC we calculated in this work.

\subsection{{\boldmath$S_\mathrm{EW}$}}  

The quantity $S_\mathrm{EW}$ was first introduced by Marciano and Sirlin in Ref.\cite{Marciano:1993sh}
as a process-independent factor that accounts for the large electroweak logarithm in the electroweak
RC \cite{Sirlin:1977sv,Sirlin:1981ie} including the $\mathcal{O}(\alpha_s)$ pQCD corrections on top
of it, as well as the resummation of the QED logs (i.e. $\delta_\mathrm{HO}^\mathrm{QED}$ in our notation).
It was often quoted schematically in the literature as~\cite{Cirigliano:2008wn,Cirigliano:2011ny}:
\begin{equation}
S_\mathrm{EW}=1+\frac{2\alpha}{\pi}\left(1-\frac{\alpha_s}{4\pi}\right)\ln\frac{M_Z}{M_\rho}+\mathcal{O}\left(\frac{\alpha\alpha_s}{\pi^2}\right)~,\label{eq:SEWana}
\end{equation}
where the $\rho$-mass appears as a low-energy scale. It is not straightforward
to infer its exact value from the expression above because some of the important components
(e.g. $\delta_\mathrm{HO}^\mathrm{QED}$) are not explicitly shown, and it is also not clear what scale one
should choose for $\alpha_s$. Fortunately, as a common consensus, the value
$S_\mathrm{EW}=1.0232(3)_\mathrm{HO}$ was always used for all practical purposes in the recent
years (see, e.g. Refs.\cite{Cirigliano:2001mk,Cirigliano:2004pv} and the FLAVIAnet global analysis,
Ref.\cite{Antonelli:2010yf}), where the central value comes from Ref.\cite{Marciano:1993sh} and
the estimated uncertainty of the QED log resummation comes from Ref.\cite{Erler:2002mv}. Notice
that although Ref.\cite{Cirigliano:2011ny} quoted a slightly different value of
$S_\mathrm{EW}=1.0223(5)$, but that number was never used in any subsequent analysis.

Now, the process-independent physics included in our $\left(\delta_{K_{e3}}\right)_\mathrm{tot}$ are not
only those described by $S_\mathrm{EW}$ but even more. For example, the most important pQCD correction
contained in $\left(\delta_{K_{e3}}\right)_{\gamma W}^{b,A}$ is calculated to $\mathcal{O}(\alpha_s^4)$
instead of just $\mathcal{O}(\alpha_s)$ in $S_\mathrm{EW}$. Therefore, it is not the most natural
choice to remove $S_\mathrm{EW}-1$ \textit{analytically} from $\left(\delta_{K_{e3}}\right)_\mathrm{tot}$
in order to compare our result with the ChPT result. Instead, it is more convenient to take the
above-mentioned numerical value of $S_\mathrm{EW}$ simply as its \textit{definition}, i.e.,
\begin{equation}
S_\mathrm{EW}-1\equiv 0.0232(3)_\mathrm{HO}~,\label{eq:SEWnum}
\end{equation}
and remove this value \textit{numerically} from $\left(\delta_{K_{e3}}\right)_\mathrm{tot}$ for
the comparison. This prescription keeps us on the same track with all the recent
literature mentioned above.

\subsection{{\boldmath$I_{Ke}^{(0)}(\lambda_i)$}}

The quantity  $I_{Ke}^{(0)}(\lambda_i)$ is formally defined as the 
``phase space integral depending on slope and curvature of the form factors $f_\pm^{K\pi}(t)$'' according to Ref.\cite{Cirigliano:2011ny}, but in practice it is treated not just as a pure QCD factor, but also contains a part of the short-distance electromagnetic effects. This can be seen in, e.g.,
Refs.\cite{Cirigliano:2001mk,Cirigliano:2004pv}: The $t$-dependence of $f_\pm^{K\pi}(t)$ at
$\mathcal{O}(p^4)$ is given by the mesonic loop functions $H_{PQ}(t)$, and we observe that in
these functions the masses of the charged mesons (e.g. $\pi^\pm$) and their neutral counterparts
(e.g. $\pi^0$) are kept distinct. Since we know that this mass splitting is partially induced
by short-distance electromagnetic effects, or more specifically, the $\mathcal{O}(e^2)$ term in the
chiral Lagrangian~\cite{Knecht:1999ag}:
\begin{equation}
\mathcal{L}_{e^2}=Ze^2F_0^4\left\langle Q_\mathrm{em}UQ_\mathrm{em}U^\dagger\right\rangle~,
\end{equation}
so the observation above implies that a part of the short-distance electromagnetic effect
proportional to $Z$ is actually assigned implicitly to $I_{Ke}^{(0)}(\lambda_i)$ through $H_{PQ}(t)$
within the ChPT framework. In our notation, this residual effect is represented exactly by the
$\bar{h}_{PQ}(t)$ terms in $\left(\delta f_{+,3}^{K\pi}(t)\right)_{e^2p^2}^\mathrm{fin}$, since the
$\bar{h}_{PQ}(t)$ functions are simply consequences from the Taylor expansion of $H_{PQ}(t)$ to
$\mathcal{O}(Z)$.

\subsection{{\boldmath$\delta_\mathrm{SU(2)}^{K\pi}$}}

The isospin-breaking correction factor $\delta_\mathrm{SU(2)}^{K\pi}$ is formally defined as\footnote{The existence of the isospin factor $C_{K^0}/C_K$ in the formula above is simply due to our choice of normalization of $f_+^{K\pi}(0)$.}:
\begin{equation}
\delta_\mathrm{SU(2)}^{K\pi}\equiv\left(\frac{C_{K^0}}{C_K}\frac{f_+^{K\pi}(0)}{f_+^{K^0\pi^-}(0)}\right)^2-1~,
\end{equation}
that is only present in $K_{l3}^+$. According to the definition above, it contains not
only the strong isospin breaking effect resulting from the $u$--$d$ mass difference, but
also the electromagnetically-induced isospin breaking. Indeed, according to Eq.(4.42) in
Ref.\cite{Cirigliano:2011ny}, one has:
\begin{equation}
\delta_\mathrm{SU(2)}^{K^\pm\pi^0}=2\sqrt{3}\left(\varepsilon^{(2)}+\varepsilon_\mathrm{S}^{(4)}+\varepsilon_\mathrm{EM}^{(4)}+...\right)~,
\end{equation}
where $\varepsilon_\mathrm{EM}^{(4)}$ originates from the electromagnetically-induced $\pi^0$--$\eta$
mixing. In our notation, this correction simply comes from $\left(\delta f_{+,3}^{K^+\pi^0}(t)
\right)_{e^2p^2}^\mathrm{fin}$ after removing the $\bar{h}_{PQ}(t)$ terms.

\subsection{{\boldmath$\delta_\mathrm{EM}^{Ke}$}}

After all the discussions above, it is now apparent that the most convenient way to discuss
$\delta_\mathrm{EM}^{Ke}$ is to simply refer it as ``the sum of all electroweak RC that are
not already contained in $S_\mathrm{EW}$, $I_{Ke}^{(0)}(\lambda_i)$ and $\delta_\mathrm{SU(2)}^{K\pi}$''.
This means
\begin{equation}
\delta_\mathrm{EM}^{Ke}=\left(\delta_{K_{e3}}\right)_\mathrm{tot}-\left(S_\mathrm{EW}-1\right)-\left(\delta_{K_{e3}}\right)_3^\mathrm{fin}
\end{equation}
in our notation, where $S_\mathrm{EW}-1$ is defined by Eq.\eqref{eq:SEWnum} as we discussed earlier.
Apart from $S_\mathrm{EW}-1$, the quantity $\left(\delta_{K_{e3}}\right)_3^\mathrm{fin}$ is also
subtracted out because its contribution is redistributed into $I_{Ke}^{(0)}(\lambda_i)$ and
$\delta_\mathrm{SU(2)}^{K\pi}$ according to the ChPT prescription, as we discussed above. In
fact, $\delta_\mathrm{EM}^{Ke}$ is also the only meaningful quantity to be compared between this
work and the existing literature, because we are taking an $\mathcal{O}(e^2p^2)$ approximation to
$\left(\delta_{K_{e3}}\right)_3^\mathrm{fin}$ and thus have made no new improvement on this term. 

\begin{table}
	\begin{centering}
		\begin{tabular}{|c|c|c|}
			\hline 
			$\delta_{\mathrm{EM}}^{Ke}$ & This work & Ref.\cite{Cirigliano:2008wn}\tabularnewline
			\hline 
			\hline 
			$K_{e3}^{0}$ & $1.16(2)_{{\color{black}\mathrm{inel}}}(1)_{\mathrm{lat}}(1^{*})_{\mathrm{NF}}(2)_{e^{2}p^{4}}\times10^{-2}$ & $0.99(19)_{e^{2}p^{4}}(11)_{\mathrm{LEC}}\times10^{-2}$\tabularnewline
			\hline 
			$K_{e3}^{+}$ & $0.21(2)_{{\color{black}\mathrm{inel}}}(1)_{\left\langle R_{K}^{2}\right\rangle }(1)_{\mathrm{lat}}(4^{*})_{\mathrm{NF}}(1)_{e^{2}p^{4}}\times10^{-2}$ & $0.10(19)_{e^{2}p^{4}}(16)_{\mathrm{LEC}}\times10^{-2}$\tabularnewline
			\hline 
		\end{tabular}
		\par\end{centering}
	        \caption{\label{tab:comparison}Comparison between the value of $\delta_\mathrm{EM}^{Ke}$
                  obtained from this work and from the ChPT calculation.}		
\end{table}

The comparison between our result of $\delta_{\mathrm{EM}}^{Ke}$ and the ChPT result is given in
Table~\ref{tab:comparison}. Let us explain all the different types of uncertainties that appear in
our new evaluation:
\begin{itemize}
\item {\color{black}inel}: This represents our conservative estimation of the effects from the {\color{black}inelastic} term
  in $\left(\delta M_2+\delta M_{\gamma W}^a\right)_\mathrm{int}$ and $\delta M_{\gamma W}^{b,V}$. See
  the discussions after Eq.\eqref{eq:GammaBorn}.
\item $\left\langle R_K^2\right\rangle$: This is the uncertainty originated from the experimental
  error of the $K^+$ charge radius (see Eq.\eqref{eq:RK2}) that enters $\left(\delta M_2
  +\delta M_{\gamma W}^a\right)_\mathrm{int}$ and $\delta M_{\gamma W}^{b,V}$ in $K_{e3}^+$. 
\item lat: This is the total lattice QCD uncertainty in the calculation of $\Box_{\gamma W}^{VA}$
  (see Eq.\eqref{eq:Boxlattice}).
\item NF: This represents our estimation of the uncertainty due to the non-forward kinematics in
  $\delta M_{\gamma W}^{b,A}$ at small loop momentum $q'$. We include an asterisk to remind the reader
  that this error estimation can be made more rigorous with an extra lattice QCD calculation, as we
  discussed at the end of Section~\ref{sec:axialbox}.
\item $e^2p^4$: This is the chiral expansion uncertainty of the non-convection term contribution
  (i.e. $2\mathfrak{Re}\left\{M_B^*M_A\right\}+|M_B|^2$, see the discussions after Eq.\eqref{eq:MAMBKp})
  in the bremsstrahlung process.
\end{itemize}

From Table~\ref{tab:comparison} we find that our results are consistent with the ChPT estimation within
the error bars, but with a significant reduction of the total uncertainty by almost an order of magnitude.
This improvement is mainly due to two reasons:
\begin{enumerate}
\item Our calculation permits a much better control of the $\mathcal{O}(e^2p^4)$ effects, which
are the main source of uncertainty in the ChPT treatment. With the new theory framework introduced
in Refs.\cite{Seng:2019lxf,Seng:2020jtz}, all the hadron physics are contained in quantities such as
$T_{K\pi}^{\mu\nu}$ and $\Gamma_{K\pi}^\mu$, from which the full convection/Born contribution can be
explicitly isolated. These contributions govern the full IR-divergent structure of the decay process,
are numerically the largest and, most importantly, do not involve any chiral expansion. The size of
the non-Born/non-convection term contributions are in general an order of magnitude smaller
(see, for example, Table~\ref{tab:bremD3} and \ref{tab:bremD4m3}), so the $\mathcal{O}(e^2p^4)$
uncertainties attached to them are even tinier. On the other hand, in the traditional ChPT
treatment one must multiply the full result by $M_K^2/\Lambda_\chi^2$ to obtain the $\mathcal{O}(e^2p^4)$
uncertainty, so it is much larger.
\item We used latest lattice QCD results to pin down $\delta M_{\gamma W}^{b,A}$, which corresponds to the
LECs $X_1$ and $\bar{X}_6^\mathrm{phys}$ in ChPT. In the existing literature, these LECs were calculated within
resonance models and were assigned a 100\% uncertainty. On the other hand, the highly-precise lattice
results of $\Box_{\gamma W}^{VA}$ would correspond exactly to $\delta M_{\gamma W}^{b,A}$ if $K$ and $\pi$ were
degenerate. We investigated the region of integration in $\delta M_{\gamma W}^{b,A}$ where this
non-degeneracy starts to take effect, and assigned a reasonable NF-uncertainty to the contribution
from this region on top of the lattice results. In the ChPT language, our treatment above simultaneously
take into account the uncertainties of the LECs themselves as well as the $\mathcal{O}(e^2p^4)$
uncertainties on top of the LEC contributions.   
\end{enumerate}

\section{\label{sec:final}Final discussions}

The $3\sigma$ discrepancy in the extraction of $V_{us}$ from $K_{\mu 2}$ and $K_{l3}$ decays has triggered
renewed interest within the particle physics community about its possible implications on the
existence of BSM physics. However, the current level of significance is not sufficient to claim a
discovery so one needs further reduction of not just the experimental errors but also the SM theory
uncertainties. Our re-analysis of the SM electroweak RC in $K_{e3}$ therefore, serves as a crucial
step along this direction. We successfully overcome the natural limitations in traditional ChPT
by adopting a new computational framework that allows for a resummation of the numerical largest
components in the RC, and also utilizing the most recent lattice QCD outcomes to reduce the
uncertainties from the non-perturbative QCD at the chiral symmetry breaking scale. Our work
reduces the existing uncertainties in the $K_{e3}$ RC by almost an order of magnitude, and finds
no large shift in the central values. This suggests that we should remove the electroweak RC
from the ``list of culprits'' responsible for the $K_{\mu 2}$--$K_{l3}$ discrepancy. 

Is it evident now that the $V_{us}$ anomaly cannot be explained by SM effects? We would say that
it is still too early to decide at this stage. Further investigations must also be made on other
SM inputs, just to mention a few:
\begin{itemize}
\item Based on the analysis of a newly-constructed ratio $R_V=\Gamma_{K_{e3}}/\Gamma_{\pi_{e3}}$,
Ref.\cite{Czarnecki:2019iwz} suggested that a shift of the lattice QCD input of $|f_+^{K^0\pi^-}(0)/
f_+^{\pi^+\pi^0}(0)|$ from its current value of $0.970(2)$ to a smaller value of $0.961(4)$ would
reconcile the $K_{\mu 2}$ and $K_{l3}$ results, and encouraged the lattice community to examine
this possibility. Lattice calculations of $|f_+^{K^0\pi^-}(0)|$ with
$N_f=2+1$~\cite{Bazavov:2012cd,Boyle:2015hfa} and
$N_f=2+1+1$~\cite{Bazavov:2013maa,Carrasco:2016kpy,Bazavov:2018kjg} in the recent years have so
far been consistent with each other, which led to the FLAG 2019 averages~\cite{FlavourLatticeAveragingGroup:2019iem}:
\begin{eqnarray}
	N_f=2+1&:&|f_+^{K^0\pi^-}(0)|=0.9677(27)\nonumber\\
	N_f=2+1+1&:& |f_+^{K^0\pi^-}(0)|=0.9706(27)~.
\end{eqnarray}
However, a new calculation by the PACS collaboration with $N_f=2+1$ returned $|f_+^{K^0\pi^-}(0)|
=0.9603(16)(^{+14}_{-4})(44)(19)(1)$ that is significantly lower than the existing
average~\cite{Kakazu:2019ltq}. This calculation utilized only one lattice spacing $a=0.085$~fm
and thus should be carefully reexamined. 
\item The quantity $I_{Kl}^{(0)}(\lambda_i)$ probes the $t$-dependence of the form factors
$\bar{f}_{+,0}(t)$. Adopting a Taylor-expansion parameterization:
\begin{equation}
  \bar{f}_{+,0}(t)=1+\lambda_{+,0}'\frac{t}{M_{\pi^+}^2}+\frac{1}{2}\lambda_{+,0}^{\prime\prime}
  \left(\frac{t}{M_{\pi^+}^2}\right)^2+...~,
\end{equation}
the parameters $\lambda_{+,0}'$ and $\lambda_+^{\prime\prime}$ are fit to the experimental distributions
of the $K_{l3}$ decays to obtain $\bar{f}_{+,0}(t)$ in the physical region of $t$. The resulting
uncertainties are 0.13\% for $I_{Ke}^{(0)}$ and 0.31\% for $I_{K\mu}^{(0)}$ (see Table~21 in
Ref.\cite{Antonelli:2009ws}), which look well under control; other forms of parameterization
were also investigated~\cite{Abouzaid:2009ry,Bernard:2006gy,Bernard:2007tk,Bernard:2009zm,Hill:2006bq}. 
However, it is known for some time that some disagreements occur in the extracted values of the
slope parameter $\lambda_0'$ of the scalar form factor from different experiments~\cite{Cirigliano:2011ny}.
Also, since $\bar{f}_{+,0}(t)$ are pure QCD quantities, their fitting to the $K_{l3}$ distributions
can only be done after removing the effects of the electroweak RC from the experimental data. Now
since we have updated the RC analysis, the fitting procedure should in principle also be updated
accordingly. Although in this paper we only present our updates of $\delta_\mathrm{EM}^{Ke}$, but
the electromagnetic corrections to the $K_{e3}$ Dalitz plots can also be derived with the same method.
\item Although the SU(2) isospin breaking correction factor $\delta_\mathrm{SU(2)}^{K\pi}$ exists only
in the $K^+$ channel by construction, its associated theory uncertainty is the largest. Upon neglecting
the electromagnetic contributions, it is given by:
\begin{equation}
  \delta_\mathrm{SU(2)}^{K\pi}=\frac{3}{2}\frac{1}{\mathcal{Q}^2}\left[\frac{M_K^2}{M_\pi^2}
    +\frac{\mathcal{Q}^2}{R}\chi_{p^4}\right]
\end{equation}
in ChPT to $\mathcal{O}(p^4)$, where $\mathcal{Q}^2\equiv(m_s^2-\hat{m}^2)/(m_d^2-m_u^2)\equiv
R(m_s/\hat{m}+1)/2$ and $\chi_{p^4}=0.219$~\cite{Gasser:1984ux}. The main uncertainties therefore
come from $\mathcal{Q}$ and $R$. For instance, disagreements are observed between the values of
$\mathcal{Q}$ and $R$ extracted from phenomenology~\cite{Colangelo:2018jxw}
\begin{eqnarray}
	\eta\rightarrow 3\pi &:&\mathcal{Q}=22.1(7),\:\:R=34.4(2.1)
\end{eqnarray}
and from lattice QCD~\cite{FlavourLatticeAveragingGroup:2019iem}
\begin{eqnarray}
	N_f=2+1&:&\mathcal{Q}=23.3(0.5),\:\:R=38.1(1.5)\nonumber\\
	N_f=2+1+1&:&\mathcal{Q}=24.0(0.8),\:\:R=40.7(2.7)
\end{eqnarray}
\end{itemize}
which must be sorted out in order to pin down the isospin breaking correction precisely.

Finally, we want to mention that we present in this work only our updates on the electroweak RC
but not a new value of $V_{us}$. A part of the reason is that we work exclusively on $K_{e3}$ and
not on $K_{\mu 3}$, given that the latter involves more sources of uncertainty (e.g. from $\delta
f_-^{K\pi}$) and will be a subject of future study. But more importantly, we realize that the physics
of kaon decay is a dynamically progressing field from where the knowledge in both experiment and
theory, including our understanding of the issues above, is being constantly updated. Therefore,
rather than quoting a new value of $V_{us}$ upon every single improvement, it is more preferable
to have a commonly agreed value that results from a collaborative work between experimentalists and
theorists based on the most updated inputs from their respective fields, similar to the FLAVIAnet
evaluation in the past decade~\cite{Antonelli:2010yf}. We hope that our research may serve as a
useful input for a possible future collaboration of such kind.

\textit{Note added:} Awaiting the review outcome of this manuscript, some of us published a new global analysis of $V_{us}$ from $K_{l3}$ based on the improvements in this work~\cite{Seng:2021nar}. The values of
$|V_{us}|$ extracted from $K_{e3}$ and $K_{\mu3}$ are currently consistent with each other within error bars, therefore we do not
see a noticeable violation of lepton flavor universality within $K_{l3}$. This requires further check from theory improvements of the $K_{\mu3}$ RC as well as future experiments. 
	
\section*{Acknowledgements} 

We thank Vincenzo Cirigliano for many inspiring discussions. This work is supported in
part by the Deutsche Forschungsgemeinschaft (DFG, German Research
Foundation) and the NSFC through the funds provided to the Sino-German Collaborative Research Center TRR110 “Symmetries and the Emergence of Structure in QCD” (DFG Project-ID 196253076 - TRR 110, NSFC Grant No. 12070131001) (U-G.M and C.Y.S), by the Alexander von Humboldt Foundation through the Humboldt
Research Fellowship (C.Y.S), by the Chinese Academy of Sciences (CAS) through a President's
International Fellowship Initiative (PIFI) (Grant No. 2018DM0034) and by the VolkswagenStiftung
(Grant No. 93562) (U-G.M), by EU Horizon 2020 research and innovation programme, STRONG-2020 project
under grant agreement No 824093 and by the German-Mexican research collaboration Grant No. 278017 (CONACyT)
and No. SP 778/4-1 (DFG) (M.G).

\begin{appendix}
	
\section{\label{sec:PS}Three- and four-body phase space in $K_{e3}$}

In this Appendix we derive the phase space formula for the $K\rightarrow \pi e^+\nu(\gamma)$ process. We start from the following master formula: suppose $A(x,y,z)$ is an arbitrary Lorentz-invariant function of the three dimensionless variables $\{x,y,z\}$ defined in Eq.\eqref{eq:xyz}, then its integration with respect to $\vec{p}'$ and $\vec{p}_e$ can be expressed as
\begin{equation}
\frac{1}{2M_K}\int\frac{d^3p'}{(2\pi)^32E'}\frac{d^3p_e}{(2\pi)^32E_e}A(x,y,z)=\frac{M_K^3}{512\pi^4}\int_{2\sqrt{r_\pi}}^\infty dz\int_{2\sqrt{r_e}}^\infty dy\int_{\alpha_-(y,z)}^{\alpha_+(y,z)}dxA(x,y,z)~,\label{eq:PSmaster}
\end{equation}
where 
\begin{equation}
\alpha_\pm(y,z)\equiv 1-y-z+r_\pi+r_e+\frac{yz}{2}\pm\frac{1}{2}\sqrt{y^2-4r_e}\sqrt{z^2-4r_\pi}~.
\end{equation}

We can apply the master formula above to derive the expressions for the $K\rightarrow \pi e^+\nu(\gamma)$ phase space. First, for $K(p)\rightarrow \pi(p') e^+(p_e)\nu(p_\nu)$, we can identify:
\begin{eqnarray}
A(x,y,z)&=&\int\frac{d^3p_\nu}{(2\pi)^32E_\nu}(2\pi)^4\delta^{(4)}(P-p_\nu)|M|^2_{K\rightarrow\pi e^+\nu}\nonumber\\
&=&\frac{2\pi}{M_K^2}\delta(x)|M|^2_{K\rightarrow\pi e^+\nu}~.
\end{eqnarray}
When plugging the expression above into Eq.\eqref{eq:PSmaster}, the $x$-integral is non-zero only when $\alpha_-(y,z)<0<\alpha_+(y,z)$, which imposes constraints on the integration region of $\{y,z\}$. Solving these inequalities gives the well-known formula:
\begin{equation}
\Gamma_{K\rightarrow\pi e^+\nu}=\frac{M_K}{256\pi^3}\int_{\mathcal{D}_3}dydz|M|^2_{K\rightarrow\pi e^+\nu}~,
\end{equation}
where the integration region $\mathcal{D}_3$ can be represented in two equivalent ways, namely:
\begin{eqnarray}
c(z)-d(z)<y<c(z)+d(z)~,&&2\sqrt{r_\pi}<z<1+r_\pi-r_e\nonumber\\
c(z)=\frac{(2-z)(1+r_e+r_\pi-z)}{2(1+r_\pi-z)}~,&&d(z)=\frac{\sqrt{z^2-4r_\pi}(1+r_\pi-r_e-z)}{2(1+r_\pi-z)}~,
\end{eqnarray}
or
\begin{eqnarray}
a(y)-b(y)<z<a(y)+b(y)~,&&2\sqrt{r_e}<y<1+r_e-r_\pi\nonumber\\
a(y)=\frac{(2-y)(1+r_\pi+r_e-y)}{2(1+r_e-y)}~,&&b(y)=\frac{\sqrt{y^2-4r_e}(1+r_e-r_\pi-y)}{2(1+r_e-y)}~.
\end{eqnarray}

\begin{figure}
	\begin{centering}
		\includegraphics[scale=0.6]{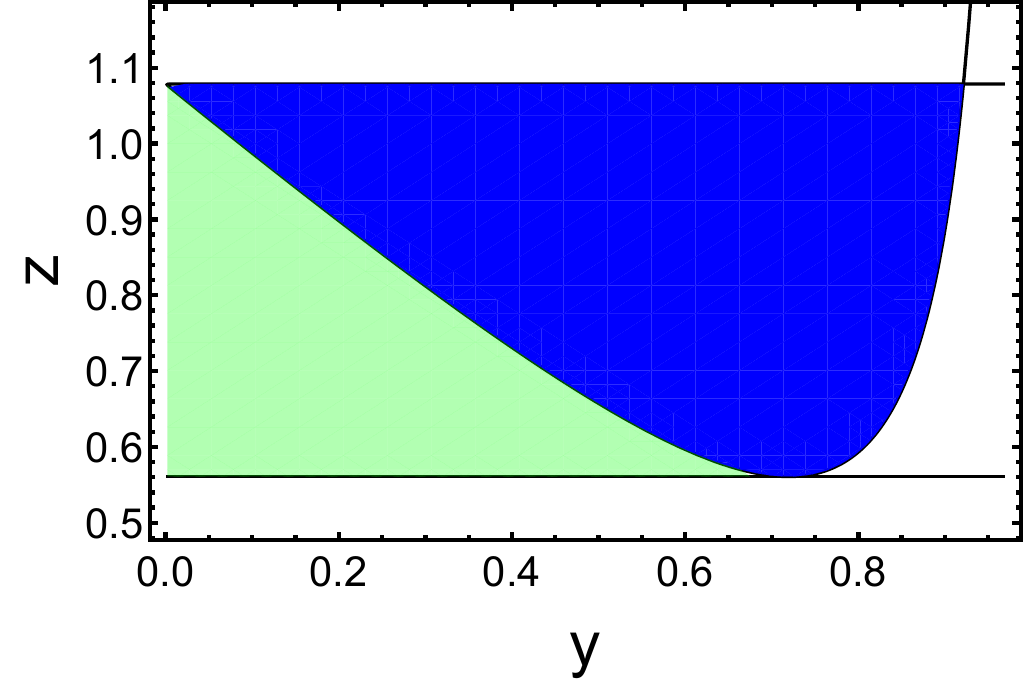}\hfill
		\par\end{centering}
	\caption{\label{fig:D3D4m3}Illustration of the $\mathcal{D}_3$ region (dark blue) and $\mathcal{D}_{4-3}$ region (light green) in $K_{e3}^0$.}
\end{figure}

Next, we discuss the phase space of $K(p)\rightarrow \pi(p') e^+(p_e)\nu(p_\nu)\gamma(k)$. In this case we can identify:
\begin{equation}
A(x,y,z)=\int\frac{d^3k}{(2\pi)^32E_k}\frac{d^3p_\nu}{(2\pi)^32E_\nu}(2\pi)^4\delta^{(4)}(P-k-p_\nu)|M|^2_{K\rightarrow \pi e^+\nu\gamma}~.
\end{equation}
Without performing the integral, one already sees that the $\delta$-function imposes the constraint $x\geq 0$ because $P^2=M_K^2x=(k+p_\nu)^2$ is just the invariant squared mass of the $\nu\gamma$ system, which cannot be negative. With that one splits the $x$-integral into two terms:
\begin{equation}
\int_{\alpha_-(y,z)}^{\alpha_+(y,z)}dx\Theta(x)=\Theta\left(\alpha_+(y,z)\right)\Theta\left(-\alpha_-(y,z)\right)\int_0^{\alpha_+(y,z)}dx+\Theta\left(\alpha_-(y,z)\right)\int_{\alpha_-(y,z)}^{\alpha_+(y,z)}dx~,
\end{equation}
and the different step functions in front of each term impose different constraints on the integration region of $\{y,z\}$. The first term requires $\alpha_-(y,z)<0<\alpha_+(y,z)$, which simply gives the $\mathcal{D}_3$ region we discussed above. Meanwhile, the second term requires $\alpha_-(y,z)>0$, and solving this inequality yields a different integration region which we may call $\mathcal{D}_{4-3}$. It can again be represented in two equivalent ways:
\begin{eqnarray}
2\sqrt{r_e}<y<c(z)-d(z)~,&&2\sqrt{r_\pi}<z<1-\sqrt{r_e}+\frac{r_\pi}{1-\sqrt{r_e}}
\end{eqnarray}
or
\begin{eqnarray}
2\sqrt{r_\pi}<z<a(y)-b(y)~,&&2\sqrt{r_e}<y<1-\sqrt{r_\pi}+\frac{r_e}{1-\sqrt{r_\pi}}~.
\end{eqnarray}
There is no overlap between the region $\mathcal{D}_3$ and $\mathcal{D}_{4-3}$ (see Fig.\ref{fig:D3D4m3}). With the above, the $K\rightarrow \pi e^+\nu\gamma$ decay rate can be written as:
\begin{eqnarray}
\Gamma_{K\rightarrow \pi e^+\nu\gamma}&=&\frac{M_K^3}{512\pi^4}\left\{\int_{\mathcal{D}_3}dydz\int_0^{\alpha_+(y,z)}dx+\int_{\mathcal{D}_{4-3}}dydz\int_{\alpha_-(y,z)}^{\alpha_+(y,z)}dx\right\}\int\frac{d^3k}{(2\pi)^32E_k}\frac{d^3p_\nu}{(2\pi)^32E_\nu}\nonumber\\
&&\times (2\pi)^4\delta^{(4)}(P-k-p_\nu)|M|^2_{K\rightarrow \pi e^+\nu\gamma}~.
\end{eqnarray}

In the study of a fully-inclusive kaon semileptonic decay rate up to $\mathcal{O}(G_F^2\alpha)$, one should add the $K\rightarrow\pi e^+\nu$ and $K\rightarrow \pi e^+\nu\gamma$ decay rate to give:
\begin{eqnarray}
\Gamma_{K\rightarrow \pi e^+\nu}+\Gamma_{K\rightarrow\pi e^+\nu\gamma}
&=&\frac{M_K}{256\pi^3}\int_{\mathcal{D}_3}dydz\left\{|M|^2_{K\rightarrow \pi e^+\nu}+\delta |M|^2_\mathrm{brem}\right\}
+\frac{M_K^3}{512\pi^4}\int_{\mathcal{D}_{4-3}}dydz\nonumber\\
&&\times\int_{\alpha_-(y,z)}^{\alpha_+(y,z)}dx\int\frac{d^3k}{(2\pi)^32E_k}\frac{d^3p_\nu}{(2\pi)^32E_\nu}(2\pi)^4\delta^{(4)}(P-k-p_\nu)|M|^2_{K\rightarrow \pi e^+\nu\gamma}~,\nonumber\\
\end{eqnarray}
where 
\begin{equation}
\delta|M|^2_\mathrm{brem}\equiv\frac{M_K^2}{2\pi}\int_0^\mathrm{\alpha_+(y,z)}dx\int\frac{d^3k}{(2\pi)^32E_k}\frac{d^3p_\nu}{(2\pi)^32E_\nu}(2\pi)^4\delta^{(4)}(P-k-p_\nu)|M|^2_{K\rightarrow \pi e^+\nu\gamma}~.\label{eq:deltaM2brem}
\end{equation}
Both $|M|^2_{K\rightarrow\pi e^+\nu}$ and $\delta |M|^2_\mathrm{brem}$ possess IR-divergences that eventually cancel other. Meanwhile, the term with the integration over the $\mathcal{D}_{4-3}$ region is by itself IR-finite.

\section{\label{sec:res}Resonances at low energy}

In this Appendix, we briefly review the basics of the resonance chiral theory that includes the $1^{--}$ and $1^{++}$ resonances as dynamical DOFs in the chiral Lagrangian~\cite{Ecker:1988te,Ecker:1989yg,Cirigliano:2006hb}. Based on this formalism we calculate the contribution of these resonance to $\delta M_2+\delta M_{\gamma W}^a$ and $\delta M_{\gamma W}^{b,V}$. 

In most of the literature on resonance chiral theory, the massive spin-1 particles are described by a totally-antisymmetric tensor field instead of a vector field~\cite{Gasser:1983yg}, so we start by introducing the formalism. First, the free Lagrangian of a (real) massive spin-1 particle is written as:
\begin{equation}
\mathcal{L}=-\frac{1}{2}\left(\partial^\lambda R_{\lambda\mu}\right)\left(\partial_\nu R^{\nu\mu}\right)+\frac{1}{4}M_R^2 R_{\mu\nu}R^{\mu\nu}~,
\end{equation} 	
where $R_{\mu\nu}$ is the antisymmetric tensor field. It satisfies the following classical equation of motion:
\begin{equation}
\partial^{\mu}\left(\partial_\lambda R^{\lambda\nu}\right)-\partial^{\nu}\left(\partial_\lambda R^{\lambda\mu}\right)+M^2_RR^{\mu\nu}=0~.
\end{equation}
The quantized field takes the form:
\begin{eqnarray}
R_{\mu\nu}(x)&=&\sum_s\int\frac{d^3k}{(2\pi)^32E_R(\vec{k})}\frac{i}{M_R}\left\{\left(k_\nu \varepsilon_\mu^s(\vec{k})-k_\mu\varepsilon_\nu^s(\vec{k})\right)\hat{a}_s(\vec{k})e^{-ik\cdot x}\right.\nonumber\\
&&\left.-\left(k_\nu \varepsilon_\mu^{s*}(\vec{k})-k_\mu\varepsilon_\nu^{s*}(\vec{k})\right)\hat{a}^\dagger_s(\vec{k})e^{ik\cdot x}\right\}~,
\end{eqnarray}
where $\varepsilon_s(\vec{k})$ is the polarization vector of the spin-1 particle that satisfies the following relations:
\begin{equation}
k\cdot\varepsilon_s(\vec{k})=0~,\:\:\:\sum_s\varepsilon_\mu^s(\vec{k})\varepsilon_\nu^{s*}(\vec{k})=-g_{\mu\nu}+\frac{k_\mu k_\nu}{M^2_R}~,
\end{equation}
and $\hat{a}^+_s(\vec{k}),\hat{a}_s(\vec{k})$ are the creation and annihilation operators. Finally, by inverting the free Lagrangian one obtains the covariant propagator of the antisymmetric tensor field:
\begin{eqnarray}
\Delta_{\mu\nu\alpha\beta}^R(p)
&=&-\frac{i}{p^2-M^2_R+i\varepsilon}\frac{1}{M^2_R}\left(g_{\mu\alpha}p_\nu p_\beta-g_{\mu\beta}p_\nu p_\alpha-g_{\nu\alpha}p_\mu p_\beta+g_{\nu\beta}p_\mu p_\alpha\right)\nonumber\\
&&+\frac{i}{M^2_R}\left(g_{\mu\alpha}g_{\nu\beta}-g_{\nu\alpha}g_{\mu\beta}\right)~.
\end{eqnarray}

We can now construct the chiral Lagrangian with dynamical vector and axial resonances. The $1^{++}$ octet resonances are represented by $V_{\mu\nu}$ which is a traceless, Hermitian matrix in the flavor space. Its chiral covariant derivative is given by:
\begin{equation}
\nabla_\lambda V_{\mu\nu}=\partial_\lambda V_{\mu\nu}+[\Gamma_\lambda,V_{\mu\nu}]~,
\end{equation}
where
\begin{equation}
\Gamma_\mu\equiv\frac{1}{2}\left\{u^\dagger\left[\partial_\mu-i(v_\mu+a_\mu)\right]u+u\left[\partial_\mu-i(v_\mu-a_\mu)\right]u^\dagger\right\}
\end{equation}
is the standard connection vector, with $v_\mu,a_\mu$ the vector and axial external sources. Similarly, the $1^{--}$ resonances are represented by the matrix $A_{\mu\nu}$. Other elementary building blocks of the ordinary ChPT. include the ``vielbein'':
\begin{equation}
u_\mu\equiv i\left\{u^\dagger\left[\partial_\mu-i(v_\mu+a_\mu)\right]u-u\left[\partial_\mu-i(v_\mu-a_\mu)\right]u^\dagger\right\}~,
\end{equation} 
and the anti-symmetric tensors $f_{R,L}^{\mu\mu}$ built from the vector and axial external sources:
\begin{equation}
f_{R,L}^{\mu\nu}\equiv\partial^\mu(v^\mu\pm a^\nu)-\partial^\nu(v^\mu\pm a^\mu)-i\left[v^\mu\pm a^\mu,v^\nu\pm a^\nu\right]~, 
\end{equation}
and finally,  $f_\pm^{\mu\nu}\equiv uf_L^{\mu\nu}u^\dagger\pm u^\dagger f_R^{\mu\nu}u$. With the above we can now write down the chiral Lagrangian with $1^{++}$ and $1^{--}$ resonances. The LO Lagrangian scales as $\mathcal{O}(p^4)$:
\begin{eqnarray}
\mathcal{L}_\mathrm{R}^{(4)}&=&-\frac{1}{2}\left\langle(\nabla^\lambda V_{\lambda\mu})(\nabla_\nu V^{\nu\mu})-\frac{1}{2}M_V^2V_{\mu\nu}V^{\mu\nu}\right\rangle-\frac{1}{2}\left\langle(\nabla^\lambda A_{\lambda\mu})(\nabla_\nu A^{\nu\mu})-\frac{1}{2}M_A^2A_{\mu\nu}A^{\mu\nu}\right\rangle\nonumber\\
&&+\frac{F_V}{2\sqrt{2}}\left\langle V_{\mu\nu}f_+^{\mu\nu}\right\rangle+\frac{iG_V}{\sqrt{2}}\left\langle V_{\mu\nu}u^{\mu}u^{\nu}\right\rangle+\frac{F_A}{2\sqrt{2}}\left\langle A_{\mu\nu}f_-^{\mu\nu}\right\rangle~,\label{eq:RChPT}
\end{eqnarray}
where $\left\langle...\right\rangle$ represents the trace over the flavor space, $M_V$ and $M_A$ are the vector and axial resonance masses in the chiral limit, while $F_V$, $F_A$ and $G_V$ are real coupling constants.

\begin{figure}
	\begin{centering}
		\includegraphics[scale=0.4]{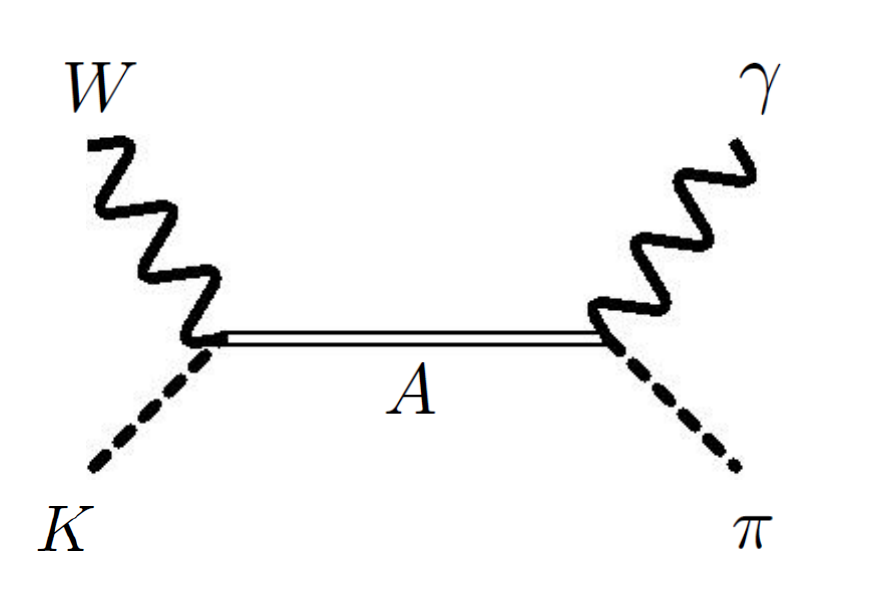}
		\includegraphics[scale=0.4]{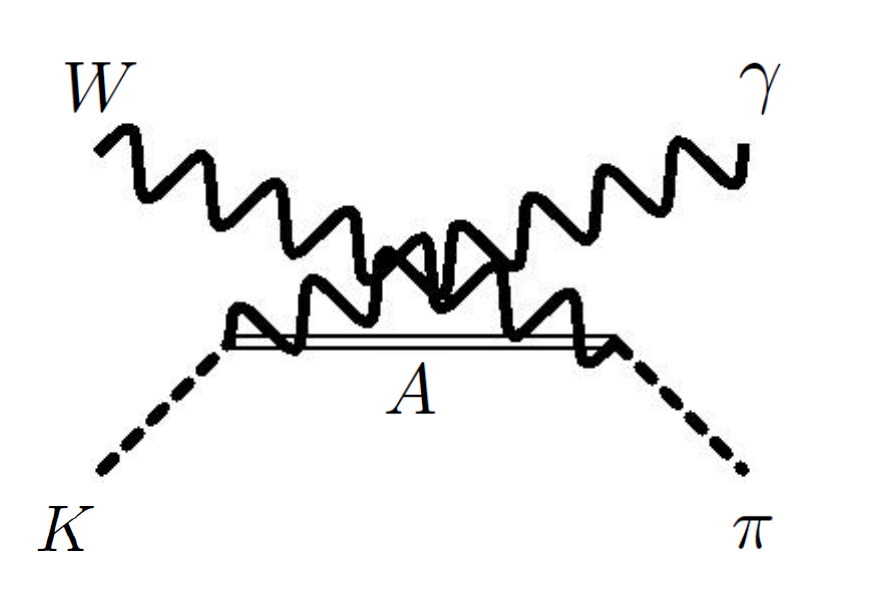}\hfill
		\par\end{centering}
	\caption{\label{fig:reslowQ}The resonance contribution to $T_{\mu\nu}^{K\pi}$ at low energy.}
\end{figure}

The leading resonance contribution to $T_{K\pi}^{\mu\nu}$ scales as $\mathcal{O}(p^4)$ and enters through the s- and u-channel diagrams as depicted in Fig.\ref{fig:reslowQ}. Since all the couplings in Eq.\eqref{eq:RChPT} have even intrinsic parity, it is evident that only the axial resonances can exist in the intermediate state. They give rise to the following expressions:
\begin{eqnarray}
\left(T^{\mu\nu}_{K^0\pi^-}(q';p',p)\right)_\mathrm{R}&=&-\frac{F_A^2}{F_0^2}V_{us}^*q'_\alpha (p-p'-q')_\beta \Delta_A^{\alpha\mu\beta\nu}(p'+q')\nonumber\\
\left(T^{\mu\nu}_{K^+\pi^0}(q';p',p)\right)_\mathrm{R}&=&\frac{F_A^2}{\sqrt{2}F_0^2}V_{us}^*q'_\alpha (p-p'-q')_\beta \Delta_A^{\alpha\mu\beta\nu}(p-q')~,\label{eq:TmunuR}
\end{eqnarray}
where $F_0$ is the pion decay constant in the chiral limit. For numerical estimation, we choose $F_A=123$~MeV, $M_A=968$~MeV following Ref.\cite{Ecker:1988te}, and $F_0\approx F_\pi=92.1$~MeV. Meanwhile, since $\Gamma^\mu_{K\pi}$ vanishes in the flavor SU(3) limit, it cannot be generated by the resonance Lagrangian in Eq.\eqref{eq:RChPT} at tree level because the latter is SU(3)-symmetric.

We then plug the expressions above into Eq.\eqref{eq:M2andMgammaWV}, \eqref{eq:deltaMWb} and evaluate the integrals. Of course, upon setting $M_W^2/(M_W^2-q^{\prime 2})\rightarrow 1$ the integrals are UV-divergent, but this is expected because the expressions above are only supposed to work at small $q'$ so the integral should be cut off at $q'\sim M_A$. As our main purpose here is just to have an order-of-magnitude estimation of the resonance contribution, we shall adopt a simple prescription as follows: we first regularize the UV-divergence using dimensional regularization, and discard the usual divergent combination $2/(4-d)-\gamma_E+\ln 4\pi$. The result is then a function of the renormalization scale $\mu$, which we vary from $M_A$ to $2M_A$ as a crude estimation of the uncertainty. With the above, we obtain the following resonance contribution to $\delta_{K_{e3}}$:
\begin{eqnarray}
\left(\delta_{K_{e3}^0}\right)_\mathrm{resonance}&=&(-0.6\pm 2.0)\times 10^{-5}\nonumber\\
\left(\delta_{K_{e3}^+}\right)_\mathrm{resonance}&=&(5.9\pm 0.8)\times 10^{-5}~.
\end{eqnarray}
They are both smaller than $10^{-4}$.

\section{\label{sec:loop}Loop functions in the convection term contributions}

In this Appendix we provide the analytic formula for the loop functions that enter the convection term contribution $\delta M_2+\delta M_{\gamma W}^a$ and $\delta M_{\gamma W}^{b,V}$.
We start by parameterizing the relevant loop integrals:
\begin{eqnarray}
\frac{(2\pi\mu)^{4-d}}{i\pi^2}\int d^dk\frac{k^\mu k^\nu}{[(p_1-k)^2-m_1^2][(p_2-k)^2-m_2^2]k^2}
&=&g^{\mu\nu}C_{00}+(p_1^\mu p_2^\nu+p_1^\nu p_2^\mu)C_{12}\nonumber\\
&&+p_1^{\mu}p_1^\nu C_{11}+p_2^\mu p_2^\nu C_{22}~,
\end{eqnarray}
\begin{equation}
\frac{1}{i\pi^2}\int d^4k\frac{k^\mu }{[(p_1-k)^2-m_1^2][(p_2-k)^2-m_2^2]k^2}=-C_1 p_1^\mu-C_2p_2^\mu~,
\end{equation}
and
\begin{equation}
\frac{1}{i\pi^2}\int d^4k\frac{1 }{[(p_1-k)^2-m_1^2][(p_2-k)^2-m_2^2][k^2-M_\gamma^2]}=C_0~.
\end{equation}
The first expression is UV-divergent and is regularized using dimensional regularization, while the third expression is IR-divergent and is regularized by a small photon mass $M_\gamma$. All the $C$s above are functions of $m_1^2=p_1^2$, $m_2^2=p_2^2$ and $v=(p_1-p_2)^2$. The analytic expressions for the $C_{ij}$ functions are as follows:
\begin{eqnarray}
C_{00}&=&\frac{1}{4}\left(\frac{2}{4-d}-\gamma_E+\ln 4\pi+\ln\frac{\mu^2}{m_1^2}+3\right)+\frac{\Lambda(v,m_1,m_2)}{4}+\frac{-m_1^2+m_2^2+v}{8v}\ln\frac{m_1^2}{m_2^2}\nonumber\\
&\equiv&\frac{1}{4}\left(\frac{2}{4-d}-\gamma_E+\ln 4\pi+\ln\frac{\mu^2}{m_1^2}+3\right)+C_{00}^\mathrm{fin}\nonumber\\
C_{11}&=&-\frac{m_1^4-2m_1^2m_2^2+m_2^4-2m_1^2v+v^2}{2v\lambda(m_1^2,m_2^2,v)}\Lambda(v,m_1,m_2)+\frac{m_1^2-m_2^2-v}{4v^2}\ln\frac{m_1^2}{m_2^2}-\frac{1}{2v}\nonumber\\
C_{22}&=&-\frac{m_1^4-2m_1^2m_2^2+m_2^4-2m_2^2v+v^2}{2v\lambda(m_1^2,m_2^2,v)}\Lambda(v,m_1,m_2)+\frac{m_1^2-m_2^2+v}{4v^2}\ln\frac{m_1^2}{m_2^2}-\frac{1}{2v}\nonumber\\
C_{12}&=&-\frac{-m_1^4+2m_1^2m_2^2-m_2^4+m_1^2v+m_2^2v}{2v\lambda(m_1^2,m_2^2,v)}\Lambda(v,m_1,m_2)-\frac{m_1^2-m_2^2}{4v^2}\ln\frac{m_1^2}{m_2^2}+\frac{1}{2v}~,\nonumber\\
\end{eqnarray}
where $\lambda(a,b,c)=a^2+b^2+c^2-2ab-2bc-2ca$ is the triangle function, and
\begin{equation}
\Lambda(v,m_1,m_2)\equiv\frac{\lambda^{1/2}(m_1^2,m_2^2,v)}{v}\ln\left(\frac{\lambda^{1/2}(m_1^2,m_2^2,v)+m_1^2+m_2^2-v}{2m_1m_2}+i\epsilon\right)~.
\end{equation}
The analytic expressions for the $C_i$ functions read:
\begin{eqnarray}
C_1&=&\frac{1}{2v}\ln\frac{m_1^2}{m_2^2}-\frac{m_1^2-m_2^2-v}{\lambda(m_1^2,m_2^2,v)}\Lambda(v,m_1,m_2)\nonumber\\
C_2&=&-\frac{1}{2v}\ln\frac{m_1^2}{m_2^2}+\frac{m_1^2-m_2^2+v}{\lambda(m_1^2,m_2^2,v)}\Lambda(v,m_1,m_2)~.
\end{eqnarray}
And finally, 
\begin{eqnarray}
C_0&=&\frac{x_v}{m_1m_2(1-x_v^2)}\left\{\ln x_v\left[-\ln\left(\frac{M_\gamma^2}{m_1m_2}\right)-\frac{1}{2}\ln x_v+2\ln(1-x_v^2)\right]-\frac{\pi^2}{6}\right.\nonumber\\
&&\left.+\mathrm{Li}_2(x_v^2)+\frac{1}{2}\ln^2\left(\frac{m_1}{m_2}\right)+\mathrm{Li}_2\left(1-x_v\frac{m_1}{m_2}\right)+\mathrm{Li}_2\left(1-x_v\frac{m_2}{m_1}\right)\right\}\nonumber\\
&\equiv&-\frac{x_v}{m_1m_2(1-x_v^2)}\ln x_v\ln\left(\frac{M_\gamma^2}{m_1m_2}\right)+C_0^\mathrm{fin}~,
\end{eqnarray}
with
\begin{equation}
x_v\equiv-\frac{1-\sqrt{1-\frac{4m_1m_2}{v-(m_1-m_2)^2}}}{1+\sqrt{1-\frac{4m_1m_2}{v-(m_1-m_2)^2}}}~.
\end{equation}

\section{\label{sec:DRIR}Dimensional regularization of the IR-divergent integral in the bremsstrahlung contribution}	

The only IR-divergent integral in bremsstrahlung process $K(p)\rightarrow \pi(p')e^+(p_e)\nu(p_\nu)\gamma(k)$ reads:
\begin{eqnarray}
I_i(y,z)&\equiv&\int_0^{\alpha_+(y,z)} dx\int\frac{d^3k}{(2\pi)^32E_k}\frac{d^3p_\nu}{(2\pi)^3 2E_\nu}(2\pi)^4\delta^{(4)}(P-k-p_\nu)\left(\frac{p_e}{p_e\cdot k}-\frac{p_i}{p_i\cdot k}\right)^2\nonumber\\
&=&2\pi\int_0^{\alpha_+(y,z)} dx\int\frac{d^3k}{(2\pi)^32E_k}\delta(M_K^2x-2k\cdot P)\left(\frac{p_e}{p_e\cdot k}-\frac{p_i}{p_i\cdot k}\right)^2,\label{eq:Ii}
\end{eqnarray}
where $i=K$ or $\pi$ (obviously, $p_K=p$ and $p_\pi=p'$). Here, we can use the single delta function in the second line to integrate out $E_k$, so the IR-divergence comes from the final integration with respect to $x$, where the integrand behaves as $x^{-1}$ at small $x$. A common prescription to regularize this IR-divergence is to introduce a non-zero photon mass such that $k^2=M_\gamma^2$. This sets a lower bound of $M_\gamma^2/M_K^2$ for the $x$-integral that regularizes the IR-divergence, but also introduces a complicated $M_\gamma$-dependence in the integrand that needs to be carefully taken into account in order to correctly reproduce all the IR-finite terms in the $M_\gamma\rightarrow 0$ limit.  

A more elegant way to deal with the IR-divergence is to use dimensional-regularization~\cite{Gastmans:1973uv,Marciano:1974tv}. With this prescription, we first generalize the three-dimensional $k$-integral to $d-1$ dimension:
\begin{eqnarray}
\frac{d^3k}{(2\pi)^3 2E_k}&\rightarrow& \mu^{4-d} \frac{d^{d-1}k}{(2\pi)^{d-1}2E_k}\nonumber\\
&=&\frac{\mu^{4-d}}{2(2\pi)^{d-1}}E_k^{d-2}\sin^{d-3}\theta_1 \sin^{d-4}\theta_2...\sin\theta_{d-3}dE_k d\theta_1 d\theta_2...d\theta_{d-3} d\theta_{d-2}~,
\end{eqnarray} 
where $0\leq \theta_1,\theta_2,...,\theta_{d-3}\leq \pi$ and $0\leq \theta_{d-2}\leq 2\pi$. 
The spatial components of $k$ are parameterized as:
\begin{eqnarray}
k_1&=&E_k\cos\theta_1\nonumber\\
k_2&=&E_k\sin\theta_1\cos\theta_2\nonumber\\
&\vdots&\nonumber\\
k_{d-2}&=&E_k\sin\theta_1\sin\theta_2...\sin\theta_{d-3}\cos\theta_{d-2}\nonumber\\
k_{d-1}&=&E_k\sin\theta_1\sin\theta_2...\sin\theta_{d-3}\sin\theta_{d-2}~.
\end{eqnarray}
With the prescription above, the IR-divergent integral over $x$ can now be simply performed:
\begin{equation}
\int_0^{\alpha_+(y,z)}dx x^{d-5}=\frac{\left(\alpha_+(y,z)\right)^{d-4}}{d-4}~,
\end{equation}
assuming $d>4$. Meanwhile, the angles can be integrated using the formula:
\begin{equation}
\int_0^\pi\sin^m\theta d\theta=\sqrt{\pi}\frac{\Gamma\left(\frac{1}{2}(m+1)\right)}{\Gamma\left(\frac{1}{2}(m+2)\right)}~.
\end{equation}
And finally, one expands the result to $\mathcal{O}\left((d-4)^0\right)$. It is also customary to switch the result back to the expression with the $M_\gamma$-regularization. For that purpose one simply performs the following matching:
\begin{equation}
\frac{2}{4-d}-\gamma_E+\ln4\pi\rightarrow \ln\frac{M_\gamma^2}{\mu^2}~.\label{eq:DRmgamma}
\end{equation}

Next, we discuss some useful tricks in the evaluation of $I_i(y,z)$ with dimensional regularization. First, the full integral can be split into three terms, with the integrand proportional to:
\begin{equation}
\frac{1}{(p_e\cdot k)^2}~,\:\:\frac{1}{(p_i\cdot k)^2}~,\:\:\frac{1}{(p_e\cdot k)(p_i\cdot k)}
\end{equation}
respectively. The integration with respect to the first term is most easily done in the $\vec{p}_e$-rest frame, while the next two terms should be done in the $\vec{p}_i$-rest frame. The following identity is also useful in performing the integration of the third term: 
\begin{eqnarray}
\int_0^{\alpha_+(y,z)}dxx^{d-5}f(d,x)&=&\int_0^{\alpha_+(y,z)}dxx^{d-5}f(d,0)+\int_0^{\alpha_+(y,z)}dxx^{d-5}\left(f(d,x)-f(d,0)\right)\nonumber\\
&=&\frac{\left(\alpha_+(y,z)\right)^{d-4}}{d-4}+\int_0^{\alpha_+(y,z)}dx\frac{1}{x}\left(f(4,x)-f(4,0)\right)+\mathcal{O}(d-4)~.\nonumber\\
\end{eqnarray}

We are now ready to write down the full result of the integral: \begin{equation}I_i(y,z)=I_i^\mathrm{IR}(y,z)+I_i^\mathrm{fin}(y,z)~,\label{eq:Iifinal}
\end{equation}
where
\begin{equation}
I_i^\mathrm{IR}(y,z)=\frac{1}{2\pi M_K^2}\left\{\left(1-\frac{1}{\beta_i(0)}\tanh^{-1}\beta_i(0)\right)\ln\left[\frac{M_K^2}{M_\gamma^2}\right]-\frac{1}{2}\ln\left[\frac{M_K^2}{m_e^2}\right]\right\}
\end{equation}
is the IR-divergent piece after switching back to the $M_\gamma$-prescription using Eq.\eqref{eq:DRmgamma}, and
\begin{eqnarray}
I_i^\mathrm{fin}(y,z)&=&\frac{1}{4\pi M_K^2}\left\{\left(1-\frac{2}{\beta_i(0)}\tanh^{-1}\beta_i(0)\right)\ln\left[\frac{M_K^2\alpha_+^2}{4P_0^2(0) }\right]+\ln\left[\frac{\alpha_+^2(y,z)}{(1-z+r_\pi-r_e)^2}\right]\right.\nonumber\\
&&-\frac{1}{\beta_i(0)}\mathrm{Li}_2\left[\frac{2\beta_i(0)}{1+\beta_i(0)}\right]+\frac{1}{\beta_i(0)}\mathrm{Li}_2\left[-\frac{2\beta_i(0)}{1-\beta_i(0)}\right]\nonumber\\
&&\left.+\frac{2}{\beta_i(0)}\mathrm{Li}_2\left[\frac{\beta_i(0)}{1+\beta_i(0)}\left(\frac{P_1(0)}{P_0(0)}+1\right)\right]-\frac{2}{\beta_i(0)}\mathrm{Li}_2\left[\frac{\beta_i(0)}{1-\beta_i(0)}\left(\frac{P_1(0)}{P_0(0)}-1\right)\right]\right\}\nonumber\\
&&-\frac{1}{2\pi M_K^2}\int_0^{\alpha_+(y,z)}dx\frac{1}{x}\left\{\frac{1}{\beta_i(x)}\ln\left[\frac{1+\beta_i(x)}{1-\beta_i(x)}\right]-\frac{1}{\beta_i(0)}\ln\left[\frac{1+\beta_i(0)}{1-\beta_i(0)}\right]\right\}\label{eq:Iifin}
\end{eqnarray}
is the IR-finite piece, with
\begin{equation}
\beta_i(x)\equiv\sqrt{1-\frac{M_i^2m_e^2}{(p_i\cdot p_e)^2}}~,\:\:P_0(x)\equiv\frac{p_i\cdot P}{M_i}~,\:\:P_1(x)\equiv\frac{1}{\beta_i(x)}\left(P_0(x)-\frac{p_e\cdot P}{p_i\cdot p_e}M_i\right)~.\label{eq:betaP}
\end{equation} 
Of course all the quantities in Eq.\eqref{eq:betaP} are functions of $\{y,z\}$ as well. Their physical meanings are apparent: $\beta_i(x)$ is the speed of the positron, $P_0(x)$ is the zeroth component of $P^\mu$, and $P_1(x)$ is the spatial component of $P^\mu$ along the direction of $\vec{p}_e$, all in the $\vec{p}_i$-rest frame. Notice that the residual, IR-finite integral in the last line of Eq.\eqref{eq:Iifin} vanishes for $i=K$, because $\beta_K(x)=\beta_K(0)$. 

The correct analytic expression for $I_K(y,z)$ and $I_\pi(y,z)$ first appeared in Ref.\cite{Ginsberg:1969jh} and Ref.\cite{Cirigliano:2004pv} respectively (notice that Ref.\cite{Ginsberg:1968pz} also attempted to calculate $I_\pi(y,z)$, but the result there is wrong even with the Errata). It is easy to check the numerical equivalence between Eq.\eqref{eq:Iifinal} and those expressions, after accounting for the difference in the overall normalization. 

\section{\label{sec:IRfin}IR-finite integrals in the bremsstrahlung contribution}

In this Appendix, we outline the general strategy to evaluate the IR-finite numerical integrations from the bremsstrahlung process, in both the $\mathcal{D}_3$ and $\mathcal{D}_{4-3}$ region. We start by providing the expressions of the relevant integrands. In  $K_{e3}^0$ we have: 
\begin{eqnarray}
|M_A|_\mathrm{res}^2&=&-e^2\left(\frac{p_e}{p_e\cdot k}-\frac{p'}{p'\cdot k}\right)^2\left\{|M_0|^2(x,y,z)-|M_0|^2(0,y,z)\right\}\nonumber\\
&&+e^2G_F^2F_\mu F_\nu^*\left(\frac{p_e}{p_e\cdot k}-\frac{p'}{p'\cdot k}\right)^2\mathrm{Tr}\left[\slashed{k}\gamma^\mu(\slashed{p}_e-m_e)\gamma^\nu(1-\gamma_5)\right]\nonumber\\
&&-e^2G_F^2F_\mu^* F_\nu\frac{1}{p_e\cdot k}\left(\frac{p_e}{p_e\cdot k}-\frac{p'}{p'\cdot k}\right)_\alpha\mathfrak{Re}\mathrm{Tr}\left[(\slashed{p}_e-m_e)\gamma^\mu(\slashed{P}-\slashed{k})\gamma^\nu\slashed{k}\gamma^\alpha(1-\gamma_5)\right]\nonumber\\
&&+e^2G_F^2F_\mu F_\nu^*\frac{1}{p_e\cdot k}\mathrm{Tr}\left[(\slashed{P}-\slashed{k})\gamma^\mu\slashed{k}\gamma^\nu(1-\gamma_5)\right]\nonumber\\
2\mathfrak{Re}\left\{M_A M_B^*\right\}&=&-2e^2G_F^2V_{us}F_\mu\left(\frac{p_e}{p_e\cdot k}-\frac{p'}{p'\cdot k}\right)_\alpha\nonumber\\
&&\times\mathfrak{Re}\mathrm{Tr}\left[(\slashed{P}-\slashed{k})\gamma^\mu(\slashed{p}_e-m_e)\left\{\frac{p^{\prime\alpha}}{p'\cdot k}\slashed{k}-\gamma^\alpha\right\}(1-\gamma_5)\right]\nonumber\\
&&-e^2G_F^2V_{us}F_\mu\frac{1}{p_e\cdot k}\mathfrak{Re}\mathrm{Tr}\left[(\slashed{P}-\slashed{k})\gamma^\mu\slashed{k}\gamma^\alpha(\slashed{p}_e-m_e)\left\{\frac{p'_\alpha}{p'\cdot k}\slashed{k}-\gamma_\alpha\right\}(1-\gamma_5)\right]\nonumber\\
|M_B|^2&=&-e^2G_F^2|V_{us}|^2\mathrm{Tr}\left[(\slashed{P}-\slashed{k})\left\{\frac{p^{\prime\mu}}{p'\cdot k}\slashed{k}-\gamma^\mu\right\}(\slashed{p}_e-m_e)\left\{\frac{p'_\mu}{p'\cdot k}\slashed{k}-\gamma_\mu\right\}(1-\gamma_5)\right]~,\nonumber\\
\end{eqnarray}
and similarly for $K_{e3}^+$, 
\begin{eqnarray}
|M_A|_\mathrm{res}^2&=&-e^2\left(\frac{p_e}{p_e\cdot k}-\frac{p}{p\cdot k}\right)^2\left\{|M_0|^2(x,y,z)-|M_0|^2(0,y,z)\right\}\nonumber\\
&&+e^2G_F^2F_\mu F_\nu^*\left(\frac{p_e}{p_e\cdot k}-\frac{p}{p\cdot k}\right)^2\mathrm{Tr}\left[\slashed{k}\gamma^\mu(\slashed{p}_e-m_e)\gamma^\nu(1-\gamma_5)\right]\nonumber\\
&&-e^2G_F^2F_\mu^* F_\nu\frac{1}{p_e\cdot k}\left(\frac{p_e}{p_e\cdot k}-\frac{p}{p\cdot k}\right)_\alpha\mathfrak{Re}\mathrm{Tr}\left[(\slashed{p}_e-m_e)\gamma^\mu(\slashed{P}-\slashed{k})\gamma^\nu\slashed{k}\gamma^\alpha(1-\gamma_5)\right]\nonumber\\
&&+e^2G_F^2F_\mu F_\nu^*\frac{1}{p_e\cdot k}\mathrm{Tr}\left[(\slashed{P}-\slashed{k})\gamma^\mu\slashed{k}\gamma^\nu(1-\gamma_5)\right]\nonumber\\
2\mathfrak{Re}\left\{M_A M_B^*\right\}&=&\sqrt{2}e^2G_F^2V_{us}F_\mu\left(\frac{p_e}{p_e\cdot k}-\frac{p}{p\cdot k}\right)_\alpha\nonumber\\
&&\times\mathfrak{Re}\mathrm{Tr}\left[(\slashed{P}-\slashed{k})\gamma^\mu(\slashed{p}_e-m_e)\left\{\frac{p^{\alpha}}{p\cdot k}\slashed{k}-\gamma^\alpha\right\}(1-\gamma_5)\right]\nonumber\\
&&+\frac{e^2G_F^2}{\sqrt{2}}V_{us}F_\mu\frac{1}{p_e\cdot k}\mathfrak{Re}\mathrm{Tr}\left[(\slashed{P}-\slashed{k})\gamma^\mu\slashed{k}\gamma^\alpha(\slashed{p}_e-m_e)\left\{\frac{p_\alpha}{p\cdot k}\slashed{k}-\gamma_\alpha\right\}(1-\gamma_5)\right]\nonumber\\
|M_B|^2&=&-\frac{e^2G_F^2}{2}|V_{us}|^2\mathrm{Tr}\left[(\slashed{P}-\slashed{k})\left\{\frac{p^{\mu}}{p\cdot k}\slashed{k}-\gamma^\mu\right\}(\slashed{p}_e-m_e)\left\{\frac{p_\mu}{p\cdot k}\slashed{k}-\gamma_\mu\right\}(1-\gamma_5)\right]~.\nonumber\\
\end{eqnarray}
In the above, we have used $F_\mu$ as a shorthand of $F_\mu^{K\pi}(p',p)$. We do not display the explicit results after taking the spinor trace, as the latter can be done with, e.g., various packages in \textit{Mathematica} such as \textit{Tracer} or \textit{Package-X}. After taking the trace, all the expressions above are functions of $\{x,y,z\}$ as well as two of the three following dot products involving $k$: $\{k\cdot p,k\cdot p',k\cdot p_e\}$ using the identity $2k\cdot (p-p'-p_e)=M_K^2x$. 

The integration can be performed with the following strategy. Take $|M_A|^2_\mathrm{res}$ in $K_{e3}^0$ as an example: we first express the squared amplitude as a finite sum:
\begin{equation}
|M_A|_\mathrm{res}^2=\sum_{m,n}c_{m,n}(x,y,z)\frac{1}{(k\cdot p')^m(k\cdot p_e)^n}~,
\end{equation}
where $-2\leq m,n \leq 2$ and $c_{m,n}(x,y,z)$ are known scalar coefficients. The $p_\nu$ and $k$-integrations return the following functions:
\begin{equation}
I_{m,n}(p_1,p_2)\equiv \frac{1}{2\pi}\int\frac{d^3k}{E_k}\frac{d^3p_\nu}{E_\nu}\frac{\delta^{(4)}(P-k-p_\nu)}{(p_1\cdot k)^m(p_2\cdot k)^n}~,
\end{equation}
of which analytic expressions are given in the Appendix of Ref.\cite{Ginsberg:1969jh} (we have checked their correctness). With this, we obtain:
\begin{equation}
\int\frac{d^3k}{(2\pi)^32E_k}\frac{d^3p_\nu}{(2\pi)^32E_\nu}(2\pi)^4\delta^{(4)}(P-k-p_\nu)|M_A|^2_\mathrm{res}=\frac{1}{8\pi}\sum_{m,n}c_{m,n}(x,y,z)I_{m,n}(p',p_e)~,
\end{equation}
where the right-hand side is now a function of $\{x,y,z\}$, so the remaining three-fold integration with respect to these variables are completely regular and can be performed numerically. The same strategy applies to the IR-finite integrals in $K_{e3}^+$, except that one should choose $1/\left\{(k\cdot p)^m(k\cdot p_e)^n\right\}$ as the basis.

\end{appendix}

\providecommand{\href}[2]{#2}\begingroup\raggedright\endgroup

\end{document}